\documentclass{jfm}

\usepackage{hyperref,xcolor}
\usepackage{natbib}
\usepackage{amsbsy}
\usepackage{psfrag}
\usepackage{amsmath}
\usepackage{soul}
\usepackage{subfig}

\usepackage{graphicx,color}
\usepackage{amssymb,latexsym}
\usepackage{url}

\def\spacce#1{\hskip #1pt}
\def\drawline#1#2{\raise 2.5pt\vbox{\hrule width #1pt height #2pt}}
\def\solid{\drawline{24}{.5}\nobreak}

\def\bdash{\hbox{\drawline{5.8}{.5}\spacce{2}}}

\def\dashed{\bdash\bdash\bdash\nobreak}

\def\bdot{\hbox{\drawline{1}{.5}\spacce{2}}}

\def\dotted{\hbox{\leaders\bdot\hskip 24pt}\nobreak}

\def\dchndot{\hbox%
{\drawline{4.6}{.5}\spacce{2}\drawline{1}{.5}\spacce{4.6}\drawline{4.6}{.5}\spacce{2}\drawline{1}{.5}}\nobreak }
\def\chndot{\hbox%
{\drawline{4.6}{.5}\spacce{2}\drawline{1}{.5}\spacce{2}\drawline{4.6}{.5}\spacce{2}\drawline{1}{.5}\spacce{2}\drawline{4.6}{.5}}\nobreak }

\def\trian{\raise 1.25pt\hbox{$\scriptstyle\triangle$}\nobreak}

\def\dtrian{\raise 1.25pt\hbox%
{$\scriptscriptstyle\bigtriangledown$}\nobreak}

\def\squar{\raise 1.25pt\hbox{$\scriptstyle\Box$}\nobreak}

\def\diamon{\raise 1.25pt\hbox{$\scriptstyle\diamond$}\nobreak}

\def\beq{\begin{equation}}
\def\eeq{\end{equation}}

\def\utau{u_\tau}

\hypersetup{
  colorlinks=true,
  linkcolor=blue,
  citecolor=blue
}

\usepackage[labelsep=period]{caption}
\renewcommand{\tablename}{\sc Table}
\renewcommand{\figurename}{\sc Figure}

\title{Characteristic scales of Townsend's wall attached eddies}
\shorttitle{Characteristic scales of Townsend's eddies} 
\shortauthor{A. Lozano-Dur\'an \& H. J. Bae } 

\author{Adri\'an Lozano-Dur\'an$^1$\corresp{\email{adrianld@stanford.edu}} and Hyunji Jane Bae$^{1,2}$ }
\affiliation{$^1$Center for Turbulence Research, Stanford University, California 94305, USA \\[\affilskip]
$^{2}$Graduate Aerospace Laboratories, California Institute of Technology, Pasadena, California, 91125}

\begin{document}

\maketitle

\begin{abstract}
%
%
\citet{Townsend1976} proposed a structural model for the logarithmic
layer (log-layer) of wall turbulence at high Reynolds numbers, where
the dominant momentum-carrying motions are organised into a
multi-scale population of eddies attached to the wall. In the attached
eddy framework, the relevant length and velocity scales of the
wall-attached eddies are the friction velocity and the distance to the
wall. In the present work, we hypothesise that the momentum-carrying
eddies are controlled by the mean momentum flux and mean shear with no
explicit reference to the distance to the wall and propose new
characteristic velocity, length, and time scales consistent with this
argument.  Our hypothesis is supported by direct numerical simulation
(DNS) of turbulent channel flows driven by non-uniform body forces and
modified mean velocity profiles, where the resulting outer-layer flow
structures are substantially altered to accommodate the new mean
momentum transfer.  The proposed scaling is further corroborated by
simulations where the no-slip wall is replaced by a Robin boundary
condition for the three velocity components, allowing for substantial
wall-normal transpiration at all lengths scales.  We show that the
outer-layer one-point statistics and spectra of this channel with
transpiration agree quantitatively with those of its wall-bounded
counterpart. The results reveal that the wall-parallel no-slip
condition is not required to recover classic wall-bounded turbulence
far from the wall and, more importantly, neither is the impermeability
condition at the wall.
\end{abstract}


\section{Introduction}

At first sight, walls appear as the most relevant constituent of
turbulence confined or limited by solid surfaces, and it seems natural
to assume that they should be the source and organising agent of
wall-bounded turbulence.  Consequently, many efforts have been devoted
to understanding the structure of turbulence in the presence of
walls. Particularly interesting is the region within the so-called
log-layer \citep{Coles1969}, where most of the dissipation resides in
the asymptotic limit of infinite Reynolds number
\citep{Marusic2013}. The seminal work by \citet{Townsend1976}
conceived the flow across the log-layer as a self-similar population
of eddies of different sizes attached to the wall and organised
according to the remaining physical quantities once viscosity is
neglected, i.e., the friction velocity and the distance to the
wall. In the present work, we propose an extension of Townsend's model
where the length and velocity scales of the momentum-carrying eddies
are controlled by the turbulent energy production rate without any
direct reference to the distance to the wall.

In addition to Townsend's attached eddy model \citep{Townsend1976} and
subsequent refinements by \citet{Perry1982}, \citet{Meneveau2013}, and
\citet{Agostini2017}, among others \citep[see][for a comprehensive
  review]{Marusic2019}, the presence of walls is key for many
low-order models and theories aiming to understand the outer-layer
dynamics. In the hairpin packet model \citep{Adrian2000}, arch-like
eddies are created at the wall and migrate away from it, although
other theories advocate for the opposite scenario of larger eddies
creating top-down effects \citep{Hunt2000}. Alternative models by
\citet{Davidson2006} and \citet{Davidson2009} do not require
wall-attached eddies but still rely on the distance to the wall as a
fundamental scaling property of the flow.  The aforementioned
proportionality of the sizes of eddies with the wall-normal distance
was originally hypothesised as an asymptotic limit at very high
Reynolds numbers and used in the classical derivation of the
logarithmic velocity profile \citep{Prandtl1925,Millikan1938} and
later iterations
\citep{Rotta1962,Coles1969,Wosnik2000,Oberlack2001,Buschmann2003}, but
it has been observed experimentally and numerically in spectra and
correlations at relatively modest Reynolds numbers in pipes
\citep{Morrison1969, Perry1975, Perry1977, Bullock1978, Kim1999,
  Guala2006, McKeon2004,Bailey2008,Hultmark2012} and in turbulent
channels and flat-plate boundary layers \citep{Tomkins2003,
  DelAlamo2004, Hoyas2006, Monty2007, Hoyas2008,
  Vallikivi2015,Chandran2017}. In this framework, the mechanism by
which eddies can ``feel'' the distance to the wall is through the no
transpiration boundary condition, i.e. impermeability.

Previous studies have also revealed that the outer flow can survive
independently of the particular configuration of the eddies closest to
the wall.  The most well-known examples are the roughness experiments
where properties of the logarithmic and outer layers remain
essentially unaltered despite the fact that roughness modifies
significantly the near-wall region \citep{Nikuradse1933,Perry1977,
  Jimenez2004, Bakken2005, Flores2006, Flores2007}.  The independence
of the outer layer with respect to the details of the near-wall region
was formulated by \cite{Townsend1976} in the context of rough walls,
and it is usually referred to as Townsend's similarity hypothesis. The
numerical study by \cite{Chung2014} assessed an idealised version of
the Townsend's similarity hypothesis by introducing slip velocities
parallel to the wall while still invoking the no transpiration
condition for the wall-normal velocity. \cite{Flores2006} showed that
the characteristics of the outer part of a channel flow remain
unchanged when perturbing the velocities at the wall. Although
transpiration at the wall was allowed, it was only for a selected set
of wavenumbers and the wall was still perceived as impermeable by most
of the flow scales. \cite{Mizuno2013} performed computations of
turbulent channels in which the wall was substituted by an off-wall
boundary condition mimicking the expected behaviour from an extended
log-layer (and hence, a wall), bypassing the buffer layer, with
relatively few deleterious effects on the flow far from the
boundaries.

Other works have pointed out that the main role of the wall is to
maintain the mean shear and that the flow motions are created at all
wall-normal distances independently of others
\citep{Lee1990,DelAlamo2006, Flores2006, Mizuno2011, Jimenez2013,
  Lozano2014b, Dong2017}.  In particular, \cite{Mizuno2011} found that
a mixing length based on the mean local shear is a better
representation than the distance to the wall as characteristic length
scale. \cite{Lee1990} revealed that streaky structures in homogeneous
turbulence at high shear rates, in which there is shear but no walls,
are similar to those observed in channels. \cite{Dong2017} further
showed that there is indeed a smooth transition between the structures
of homogeneous shear turbulence and the attached eddies of
wall-bounded flows. Consistent with the previous idea, it was shown by
\cite{Tuerke2013} that even minor artificial changes in the mean shear
can lead to significant modifications of the turbulence structure.  A
growing body of evidence also indicates that the generation of the
log-layer eddies originates from the interaction with the mean shear
via linear lift-up effect
\citep{DelAlamo2006,Hwang2010,Moarref2013,Alizard2015}, and
\cite{Hwang2016} have further shown that attached log-layer eddies
follow an individual self-sustaining process independently of the
details of the surrounding scales.

Among the previous studies, the underlying tentative consensus is that
most of the energy- and momentum-carrying eddies of the log-layer are
attached, in the sense that they span the full distance from the wall
to their maximum height \citep{Townsend1961}, and that they evolve
independently of the details of the buffer layer dynamics or even
other scales.  Recent surveys have been given by \cite{Adrian2007,
  Smits2011, Jimenez2012} and \cite{Jimenez2018}. However, the
wall-normal distance remains a foundational element for most theories
and models for wall turbulence, and the identification of the
characteristic scales controlling momentum-carrying eddies consistent
with a wall-independent formulation calls for further investigation.
In the present work, we propose characteristic velocity, length, and
time scales of the momentum-carrying eddies (i.e., Townsend's eddies)
by assuming that they are predominantly controlled by the mean
momentum flux and mean shear. A preliminary version of the work can be
found in \citet{Lozano2019}.

The paper is organised as follows. The new characteristic scales are
presented in \S \ref{sec:scales}.  The proposed scaling is assessed in
turbulent channel flows with modified mean momentum flux in \S
\ref{sec:results:flux}. The results show that the new scaling is
satisfactory for the range of cases investigated. In \S
\ref{sec:results:slip}, we introduce turbulent channels where the
no-slip walls are replaced by Robin boundary conditions for the three
velocity components. In this new set-up, transpiration is allowed at
the boundary, blurring the wall-normal reference imposed by the
impermeable wall.  We show that the one-point statistics, spectra, and
three-dimensional structure of the eddies responsible for the momentum
transfer of a classic no-slip channel are identical to those obtained
by replacing the wall by a Robin boundary condition. Finally,
concluding remarks are made in \S \ref{sec:conclusions}.
%

\section{Scales controlling momentum-carrying eddies}\label{sec:scales}

We will consider a statistically steady, wall-bounded turbulent flow
confined between two parallel walls where $u_i$ with $i=1,2,3$ are the
streamwise, wall-normal, and spanwise velocities, respectively. The
pressure is denoted by $p$. The three spatial directions are $x_i$
with $i=1,2,3$, and the walls are located at $x_2=0$ and $x_2=2h$
where $h$ is the channel half-height. The fluid is incompressible with
density $\rho$ and kinematic viscosity $\nu$. We further assume that
the flow is homogeneous in the streamwise and spanwise directions.

First, we briefly revisit the classic scaling by \citet{Townsend1976}.
The traditional argument for the characteristic velocity of eddies
transporting tangential Reynolds stress is that their associated
turbulence intensities equilibrate to comply with the mean integrated
momentum balance
\begin{equation}
\langle u_1 u_2 \rangle \approx u_\tau^2 \left( \frac{x_2}{h} -1 \right),
\label{eq:utau}
\end{equation}
where the viscous effects have been neglected, $\langle\cdot\rangle$
denotes averaging in the homogeneous directions and time, and $u_\tau$
is the so-called friction velocity defined as $u_\tau =
\sqrt{\tau_w/\rho}$, where $\tau_w$ is total wall shear stress. For a
no-slip wall, the friction velocity reduces to $u_\tau =\sqrt{\nu
  \partial\langle u_1\rangle/\partial x_2}|_{x_2=0}$.  Hence, the
relevant velocity scale at all wall-normal distances is identified as
$u_\tau$.  Regarding the characteristic length scale, the classic
theory states that the log-layer motions are too large to be affected
by viscosity but small compared to the most restrictive boundary layer
limit $\mathcal{O}(h)$. It is argued then that the most meaningful
length scale the log-layer eddies can be influenced by is the distance
to the wall.

\begin{figure}
\begin{center}
 \vspace{0.5cm}
 \includegraphics[width=0.90\textwidth]{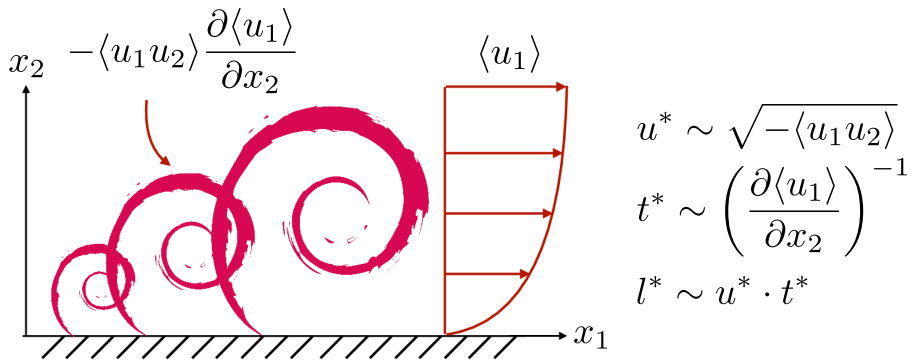}
\end{center}
\caption{ Sketch of wall-attached momentum-carrying eddies of
  different sizes in a turbulent boundary layer controlled by the mean
  production rate of turbulent kinetic energy, $-\langle u_1 u_2
  \rangle \partial \langle u_1 \rangle/\partial x_2$, and proposed
  velocity $u^*$, time $t^*$, and length $l^*$
  scales. \label{fig:model}}
\end{figure}

The hypothesis under consideration in the present work is that the
wall is not the organising element of the momentum-carrying eddies,
whose intensities and sizes are controlled instead by the mean
production rate of turbulent kinetic energy, i.e., by the mean
momentum flux $-\langle u_1 u_2 \rangle$ and associated mean shear
$\partial \langle u_1 \rangle /\partial x_2$. The proposed
characteristic length $l^*$, time $t^*$, and velocity $u^*$ scales of
wall-attached eddies are sketched in figure \ref{fig:model}.
%

The characteristic velocity scale proposed in the present work is
dictated by the momentum flux
\begin{equation}
u^* \equiv \sqrt{-\langle u_1 u_2 \rangle}. 
\label{eq:u_star}
\end{equation}
For direct comparisons with the classic scaling $u_\tau$, we can
define an alternative characteristic velocity scale by analogy with
(\ref{eq:utau}) as $u^\star \equiv \sqrt{ -\langle u_1 u_2 \rangle/(
  1-x_2/h ) }$ \citep{Tuerke2013}.  Note that the factor
$\sqrt{1-x_2/h}$ in the definition of $u^\star$ is introduced only for
convenience and such that $u^\star$ collapses to $u_\tau$ for a
channel flow driven by a constant mean pressure gradient.

The scenario proposed for the characteristic time and length scales of
wall-attached eddies also differs from the classic theory. Let us
consider a momentum-carrying eddy of size $l^* \sim u^* \cdot t^*$
controlled by the injection of energy from the mean shear and,
therefore, with characteristic lifespan
\begin{equation}
t^*\equiv \left( \frac{\partial \langle u_1 \rangle }{\partial x_2} \right)^{-1}.
\label{eq:t_star}
\end{equation}
The scaling in (\ref{eq:t_star}) can also be interpreted as the
average time for the eddies to extract energy from the mean
shear. Considering the scaling proposed for $u^*$ in
(\ref{eq:u_star}), the characteristic length scale is
\begin{equation}
l^* \equiv u^* \cdot t^* = \sqrt{ -\langle u_1 u_2 \rangle} 
\left( \frac{\partial \langle u_1 \rangle }{\partial x_2}
\right )^{-1}.
\label{eq:l_star}
\end{equation}
Equation (\ref{eq:l_star}) can also be obtained by assuming that the
momentum-carrying eddies are controlled by the mean production rate of
turbulent kinetic energy, $P = -\langle u_1 u_2 \rangle \partial
\langle u_1 \rangle /\partial x_2 \sim u^{*3}/l^*$, which together
with (\ref{eq:u_star}) yields an expression identical to
(\ref{eq:l_star}).

For the particular case of a plane channel flow with no-slip walls and
constant mean pressure gradient we recover $u^\star = u_\tau$ away
from the wall. Moreover, at high Reynolds numbers, the characteristic
length $l^*$ in the log-region of the flow ($x_2 \lesssim 0.2 h$) with
mean shear $\partial \langle u_1\rangle/\partial x_2 = u_\tau/(\kappa
x_2)$ reduces to
\begin{equation}
l^* = u_\tau \sqrt{1-x_2/h} \left( \frac{u_\tau}{\kappa x_2} \right )^{-1} 
= \kappa x_2 \sqrt{1-x_2/h}  \approx \kappa x_2,
\end{equation}
which is proportional to the distance to the wall as commonly
discussed in the literature. Therefore, the extension of the
characteristic scales proposed above collapses to Townsend's model for
a canonical case. It is important to remark that despite the fact that
the velocity and length scales specified by $u^\star$ and $l^*$
coincide with their classic counterparts $u_\tau$ and $x_2$ for the
traditional channel flow, the former are conceptually different as
they remain agnostic to the location of the wall.
%

%

\section{Turbulent channel driven by $x_2$-dependent body force}\label{sec:results:flux}

\subsection{Numerical scheme and computational domain}\label{subsec:numerical:code}

We perform a set of DNS of plane turbulent channel flows by solving
the incompressible Navier-Stokes equations with a staggered,
second-order, finite difference \citep{Orlandi2000} and a
fractional-step method \citep{Kim1985} with a third-order Runge-Kutta
time-advancing scheme \citep{Wray1990}. Periodic boundary conditions
are imposed in the streamwise and spanwise directions, and no-slip at
the walls. The code has been validated in previous studies in
turbulent channel flows \citep{Lozano2016_Brief,Bae2018b,Bae2018} and
flat-plate boundary layers \citep{Lozano2018}.

Wall units are denoted by the superscript $+$ and defined in terms of
the kinematic viscosity $\nu$ and friction velocity at the wall
$u_\tau$.  Accordingly, the friction Reynolds number is
$Re_\tau=u_\tau h/\nu$.  Velocities normalised by $u^*$ and $u^\star$
are denoted by the superscript $*$ and $\star$,
respectively. Fluctuating quantities with respect to the mean are
represented by $(\cdot)'$.  The computational domain for the present
simulations is $2\pi h\times 2h\times \pi h$ in the streamwise,
wall-normal, and spanwise directions, respectively. It has been shown
that this domain size is large enough to accurately predict one-point
turbulent statistics for $Re_\tau$ up to 4200 \citep{Lozano2014a}.

We compare our results with DNS data from \cite{DelAlamo2003} and
\cite{Hoyas2006} at $Re_\tau\approx 550, 950$, and $2000$, which are
labelled as NS550, NS950, and NS2000, respectively. The three cases
have computational domains equal to $8\pi h \times 2h \times 3\pi h$
in the streamwise, wall-normal, and spanwise directions, respectively.

\subsection{Numerical experiments of channels driven by $x_2$-dependent body force}\label{subsec:numerical:flux}

We devise two sets of conceptual numerical experiments to unravel the
characteristic scales of the outer-layer motions.
The first set of experiments is a channel flow with no-slip walls
driven by a $x_2$-dependent body force per unit mass of the form
\begin{equation}
f_1 = \frac{u_\tau^2}{h} \left[ 1 + \epsilon \left(2 x_2/h -  x_2^2 / h^2 \right) - 2/3\epsilon \right], \
f_2 = 0, \ f_3 = 0,
\label{eq:new_dP}
\end{equation}
where $f_i$ is the component of the force in the $i$-th direction, and
$\epsilon$ is a non-dimensional adjustable parameter. Equation
(\ref{eq:new_dP}) is such that $Re_\tau$ remains unchanged with
$\epsilon$. For $\epsilon=0$, we recover the constant body force
typically used to drive the channel, $f_1 = u_\tau^2/h$.  The goal of
(\ref{eq:new_dP}) is to alter the natural balance between eddies,
which are forced to readjust their intensities to accommodate the new
momentum flux. Two cases are considered at $Re_\tau \approx 550$:
$\epsilon = 4$, labelled as NS550-p, and $\epsilon = -2$, labelled as
NS550-n, where NS denotes no-slip boundary condition at the walls.
Note that case NS550 corresponds to $\epsilon = 0$.

The second set of experiments intends to clarify the characteristic
length scales of the active energy-containing eddies. The change in
$u^*$ and $t^*$ from NS550-p and NS550-n is not significant enough to
assess conclusively the scaling proposed in (\ref{eq:l_star}). For
that reason, two new simulations, NS550-s1 and NS550-s2, are
considered by prescribing a synthetic mean velocity profile of the
form
\begin{equation}
\frac{\langle u_1 \rangle}{u_{\mathrm{ref}}} = 
   \frac{\alpha+2}{\alpha+1}\left[ 1-(x_2/h-1)^{\alpha+1} \right] 
+  \frac{\beta+2}{\beta+1}\left[1-(x_2/h-1)^{\beta+1} \right],
\label{eq:new_mean}
\end{equation}
with $(\alpha,\beta)=(41,11)$ and $(3,3)$ for NS550-s1 and NS550-s2,
respectively. The parameter $u_{\mathrm{ref}}$ was adjusted to achieve
$Re_\tau \approx 550$. The profiles from (\ref{eq:new_mean}) are
purposely tailored to create distinguishable $l^*$ with values equal
to $0.06h$, $0.03h$, and $0.04h$ at $x_2=0.10h$, for cases NS550,
NS550-s1, and NS550-s2, respectively. These last two simulations are
similar to channel flows driven by $x_2$-dependent body forces
discussed above, in the sense that prescribing the mean velocity
profile is equivalent to imposing a $x_2$-dependent (and
time-dependent) forcing as in (\ref{eq:new_dP}). Simulations of
turbulent channels with prescribed velocity profiles can also be found
in \cite{Tuerke2013}. All the cases above are designed such that
$u_\tau$ and $x_2$ remain unchanged but do not coincide with the
scaling proposed by $u^\star$ and $l^*$, in contrast with the
traditional channel flow, where $u^\star\approx u_\tau$ and
$l^*\approx \kappa x_2$. This will allow us to assess the validity of
each scaling.  The list of cases is summarised in table
\ref{tab:cases:flux}.  All the simulations were run for at least 10
eddy turnover times (defined as $h/u_\tau$) after transients.
%
\begin{table}
\begin{center}
\setlength{\tabcolsep}{12pt}
\begin{tabular}{l c c c c c c c c}
Case       &  $\Rey_\tau$ & $\Delta x_1^+$  & $\Delta x^+_{2,\mathrm{min}/\mathrm{max}}$  & $\Delta x_3^+$ & Driven by \\
\hline
\hline                                              
NS550-p      & 546          & 6.7   & 0.2/9.9  & 3.3   &   $f_1$ with $\epsilon=4$ \\
NS550-n      & 546          & 6.7   & 0.2/9.9  & 3.3   &   $f_1$ with $\epsilon=-2$\\
NS550-s1     & 531          & 6.5   & 0.2/9.7  & 3.2   &   prescribed $\langle u_1 \rangle$  \\
NS550-s2     & 546          & 6.7   & 0.2/9.9  & 3.3   &   prescribed $\langle u_1 \rangle$  \\
\hline
\end{tabular}
\end{center}
\caption{Tabulated list of cases for \S \ref{sec:results:flux}. The
  numerical experiments are labelled following the convention
  NS[$\Rey_\tau$]-[specific case], where NS denotes channel with
  no-slip walls.  $\Delta x_1$, $\Delta x_2$ and $\Delta x_3$ are the
  streamwise, wall-normal, and spanwise grid resolutions,
  respectively. The last column shows the method employed to drive the
  channel flow. See text for details.}
\label{tab:cases:flux}
\end{table}

\subsection{Assessment of characteristic velocity and length scales}\label{subsec:results:flux}

\begin{figure}
\begin{center}
 \vspace{0.1cm}
 \psfrag{X}{$x_2/h$}\psfrag{Y}[cb]{$-\langle u_1^+ u_2^+ \rangle + \partial \langle u_1^+ \rangle /\partial x_2^+$}
 \subfloat[]{ \includegraphics[width=0.46\textwidth]{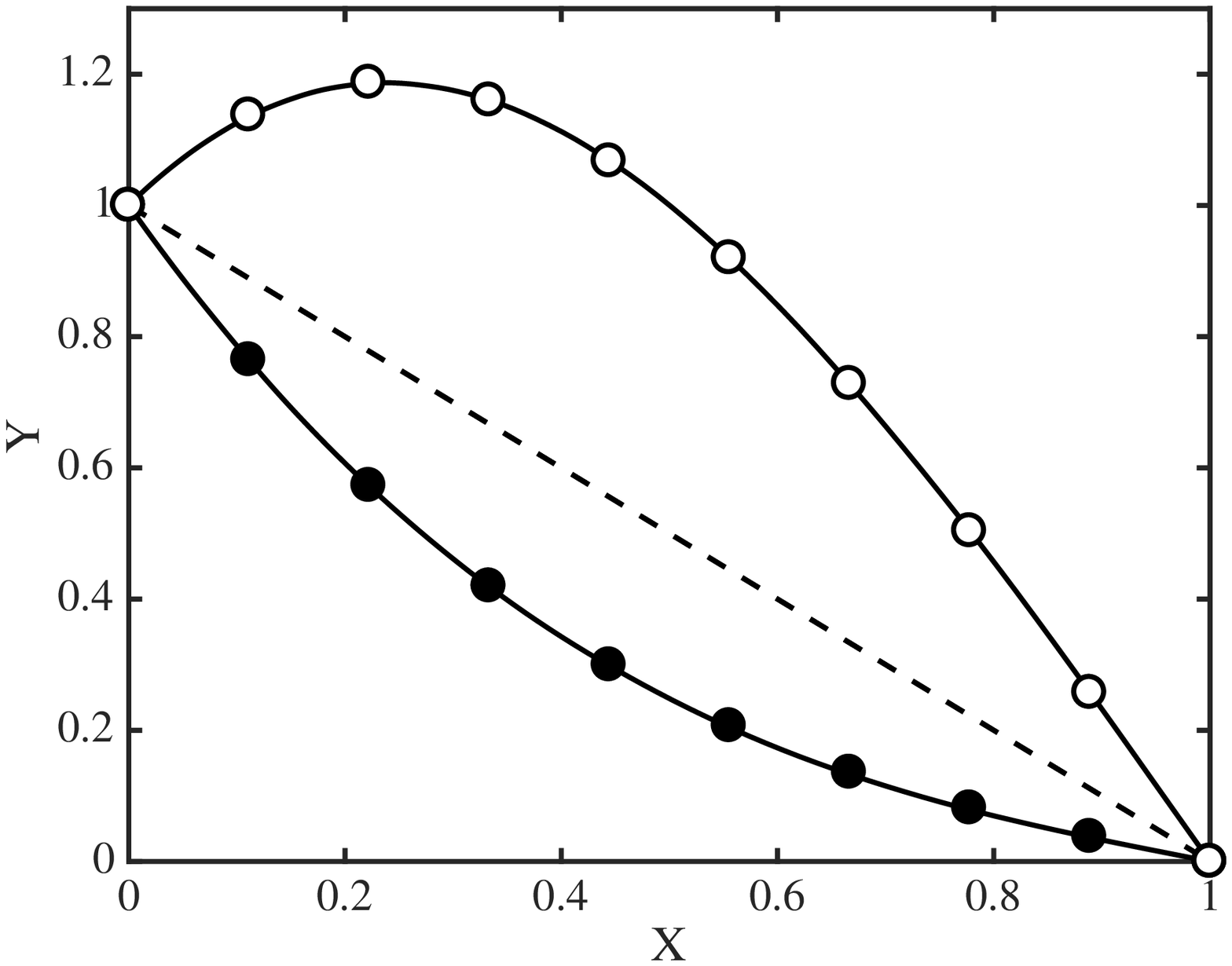} }
 \hspace{0.2cm}
 \psfrag{X}{$x_2^+$}\psfrag{Y}{$\langle u_1^+ \rangle$}
 \subfloat[]{ \includegraphics[width=0.46\textwidth]{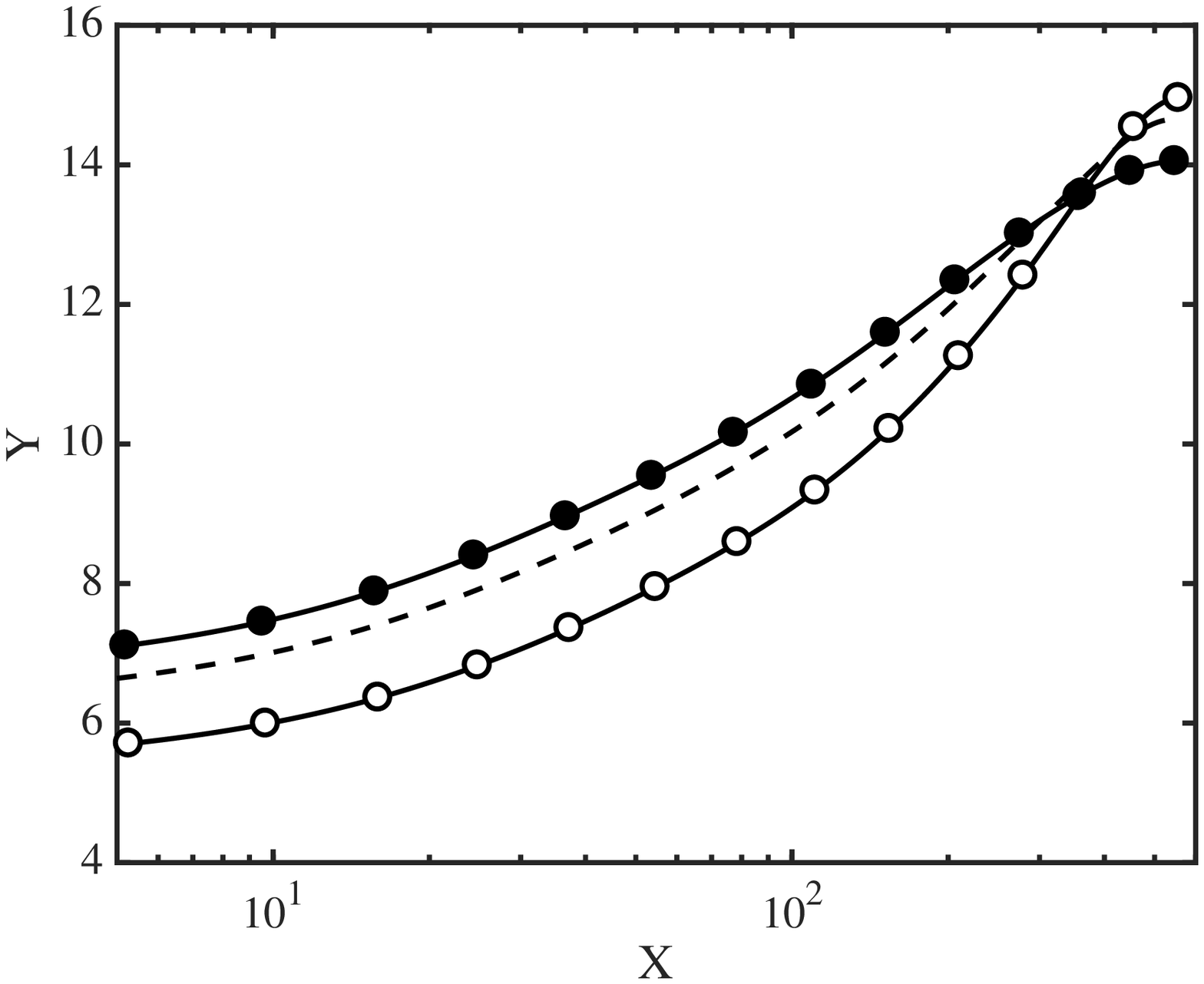} }
 \end{center}
%
%
 \begin{center}  
   \psfrag{X}{$x_2/h$}\psfrag{Y}{$\langle {u_i'}^{2+} \rangle^{1/2}$}
   \subfloat[]{ \includegraphics[width=0.46\textwidth]{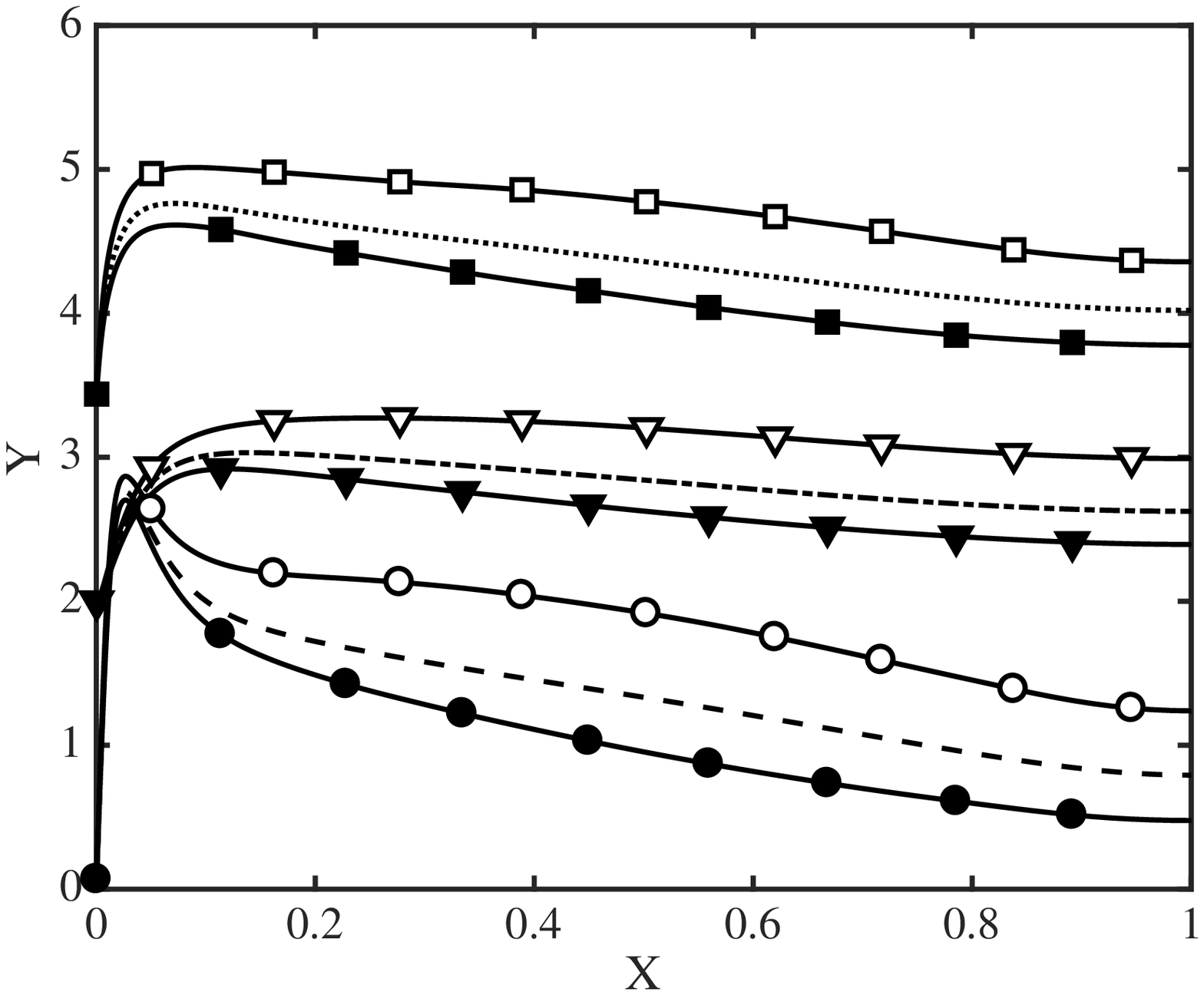} }
   \hspace{0.2cm}
   \psfrag{X}{$x_2/h$}\psfrag{Y}{$\langle {u_i'}^{2\star} \rangle^{1/2}$}
   \subfloat[]{ \includegraphics[width=0.46\textwidth]{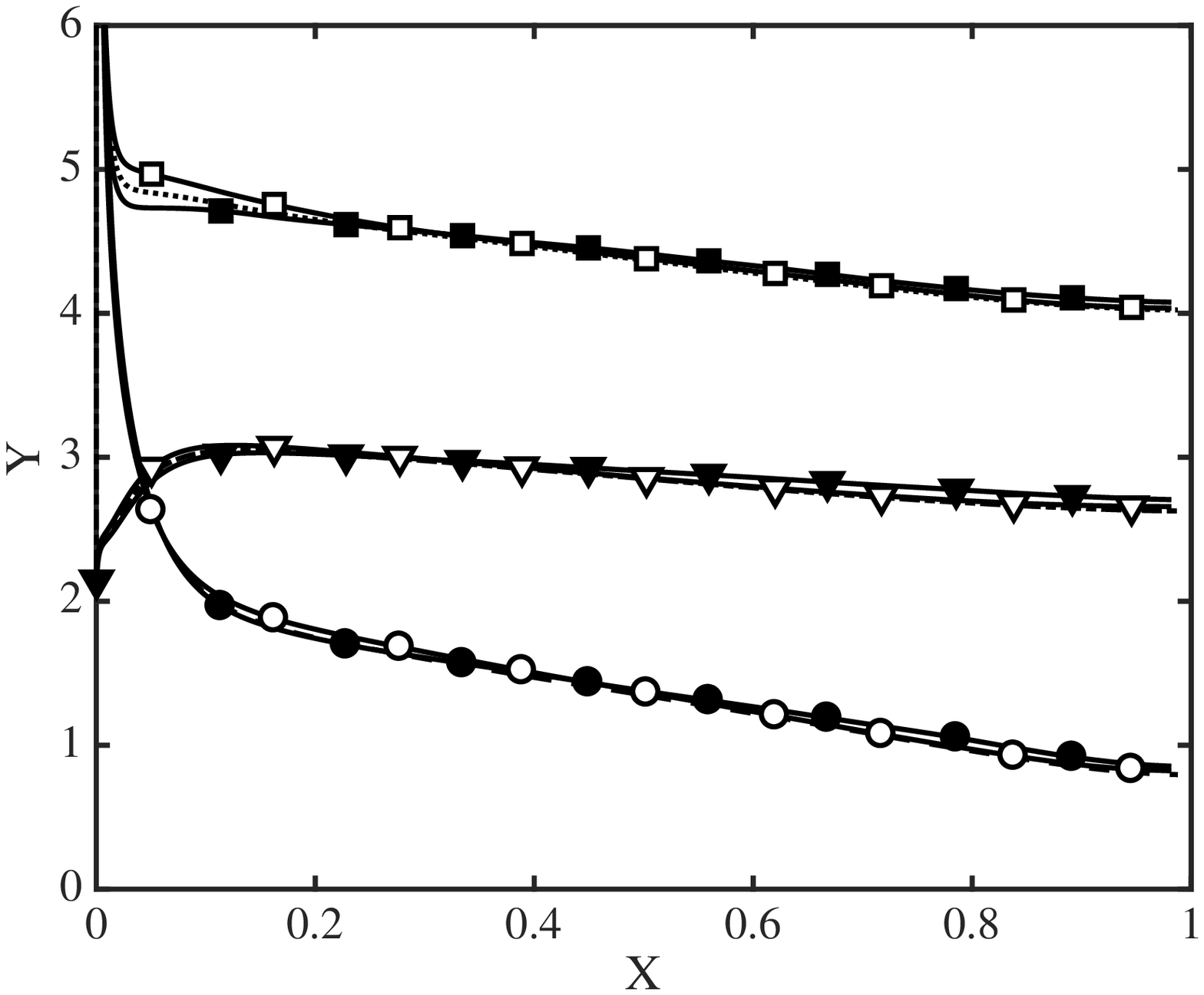} }
 \end{center}
\caption{ Mean (a) total tangential Reynolds stress and (b) streamwise
  velocity profile for NS550 (\dashed), NS550-p ($\circ$), and NS550-n
  (\textbullet). Panels (c) and (d) contain the streamwise (circles
  and \dashed), wall-normal (triangles and \chndot), and spanwise
  (squares and \dotted) root-mean-squared velocity fluctuations scaled
  with (c) $u_\tau$ and (d) $u^\star$. Symbols are for NS550-p (open)
  and NS550-n (closed). Lines without symbols are for NS550.  For
  clarity, the profiles for the wall-normal and spanwise
  root-mean-squared velocity fluctuations are shifted vertically by
  2.0 and 3.4 wall units. \label{fig:rms_eps}}
\end{figure}

We examine scaling (\ref{eq:u_star}) in turbulent channel flows driven
by (\ref{eq:new_dP}).  The imposed $x_2$-dependent body force breaks
the global velocity scale with $u_\tau$, and the new balance for the
mean momentum flux requires that
\begin{equation}
\langle u_1 u_2 \rangle \approx 
\int_0^{x_2} f_1(x'_2) \mathrm{d}x'_2.
\label{eq:int_dp}
\end{equation}
The total stresses consistent with (\ref{eq:int_dp}) for cases NS550-p
and NS550-n are shown in figure \ref{fig:rms_eps}(a).  Changes in the
momentum flux propagate to the mean velocity profile as dictated by
the integrated streamwise mean momentum equation, and the resulting
profiles are shown in figure \ref{fig:rms_eps}(b).
%

The three root-mean-squared (r.m.s.) fluctuating velocities for NS550,
NS550-p, and NS550-n are reported in figure \ref{fig:rms_eps}(c). The
pronounced lack of collapse among the three cases exposes the
unsatisfactory scaling with $u_\tau$. Conversely, when the r.m.s.
fluctuating velocities are scaled with $u^\star$, which can be
analytically evaluated for cases NS550-p and NS550-n, the agreement is
excellent (figure \ref{fig:rms_eps}d). Appendix \ref{sec:appendixB}
shows results using $u^*$, which does not have any explicit
functional dependence on $x_2$. Note that the argument above holds for
Townsend's active motions, i.e., those responsible for the mean
momentum transfer, and that the inactive motions are not expected to
scale with $u^\star$ (or $u^*$) but with the bulk velocity or a mixed
scale as suggested in previous works
\citep{Zagarola1998,DeGraaff2000,DelAlamo2004,Morrison2004}.

\begin{figure}
 \begin{center}  
   \vspace{0.1cm}
   \psfrag{X}{$x_2^+$}\psfrag{Y}{$\langle u_1^+ \rangle$}
   \subfloat[]{ \includegraphics[width=0.46\textwidth]{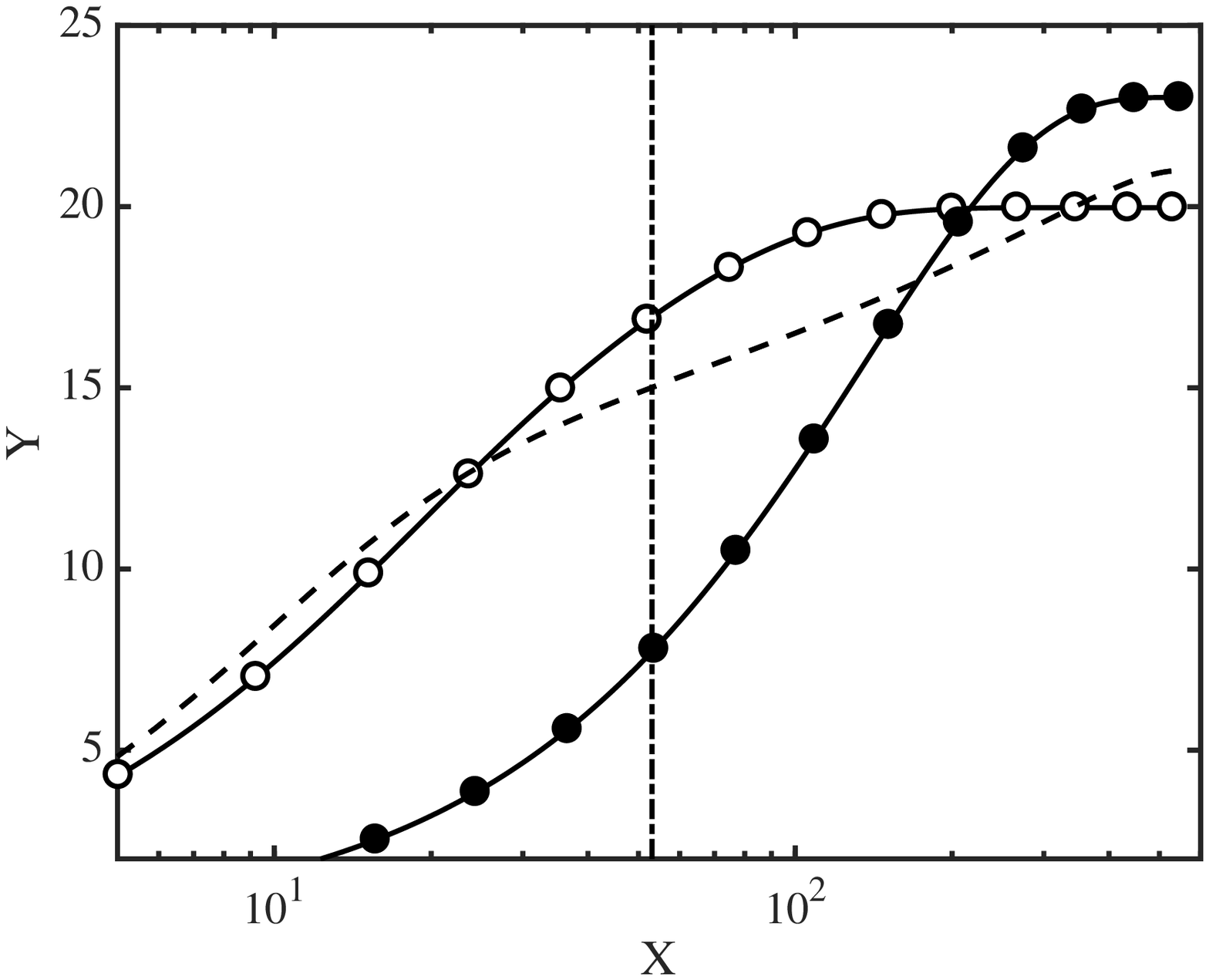} }
   \hspace{0.2cm}
   \psfrag{X}{$x_2/h$}\psfrag{Y}[cb]{$-\langle u_1^+ u_2^+ \rangle$}
   \subfloat[]{ \includegraphics[width=0.46\textwidth]{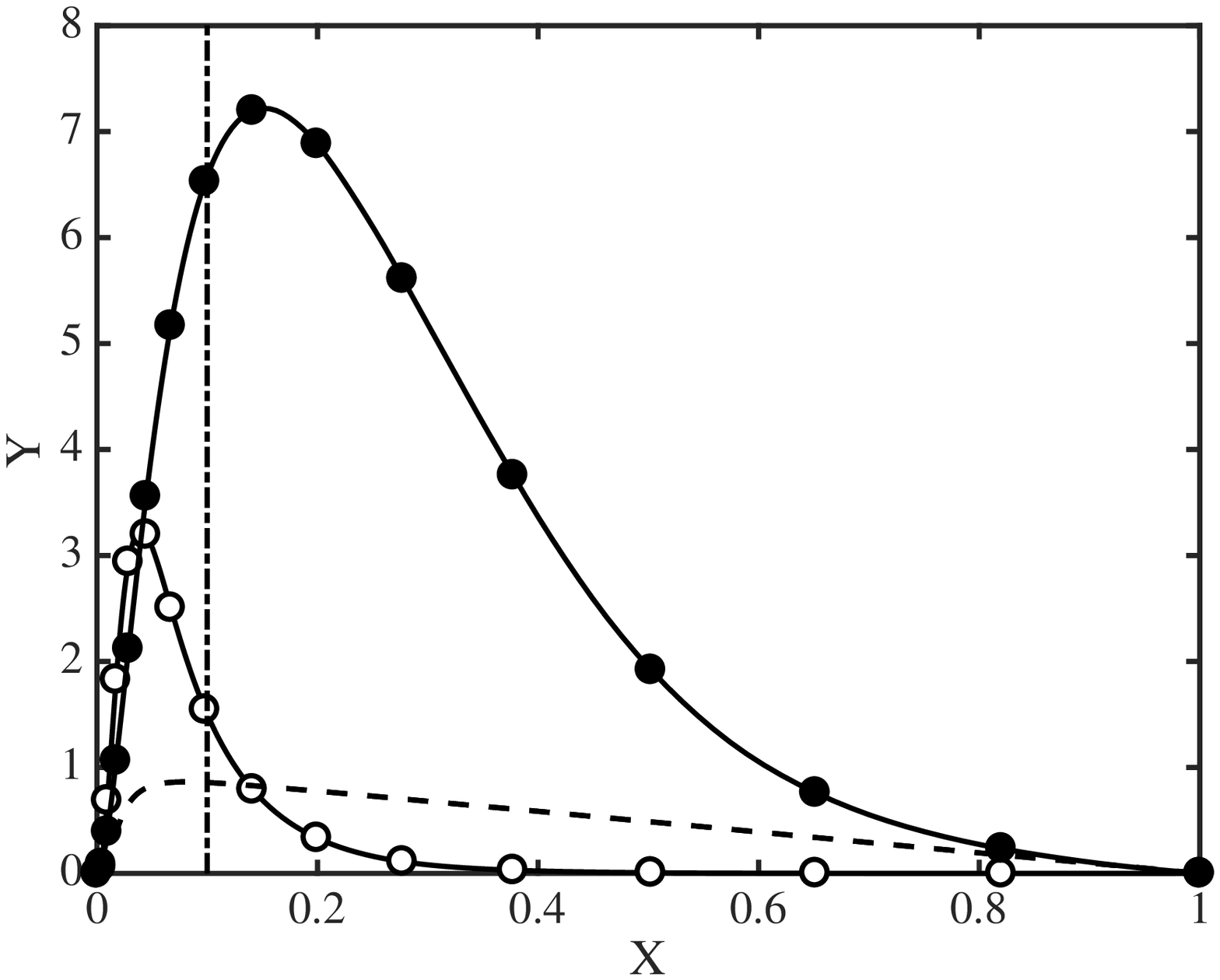} }
 \end{center}
\caption{ Mean (a) streamwise velocity profile and (b) tangential
  Reynolds stress for NS550 (\dashed), NS550-s1 ($\circ$), and
  NS550-s2 (\textbullet). The dash-dotted line is $x_2 = 0.1h$.
\label{fig:Umean_fixed}}
\end{figure}
\begin{figure}
 \vspace{0.1cm}
 \begin{center}  
   \includegraphics[width=0.95\textwidth]{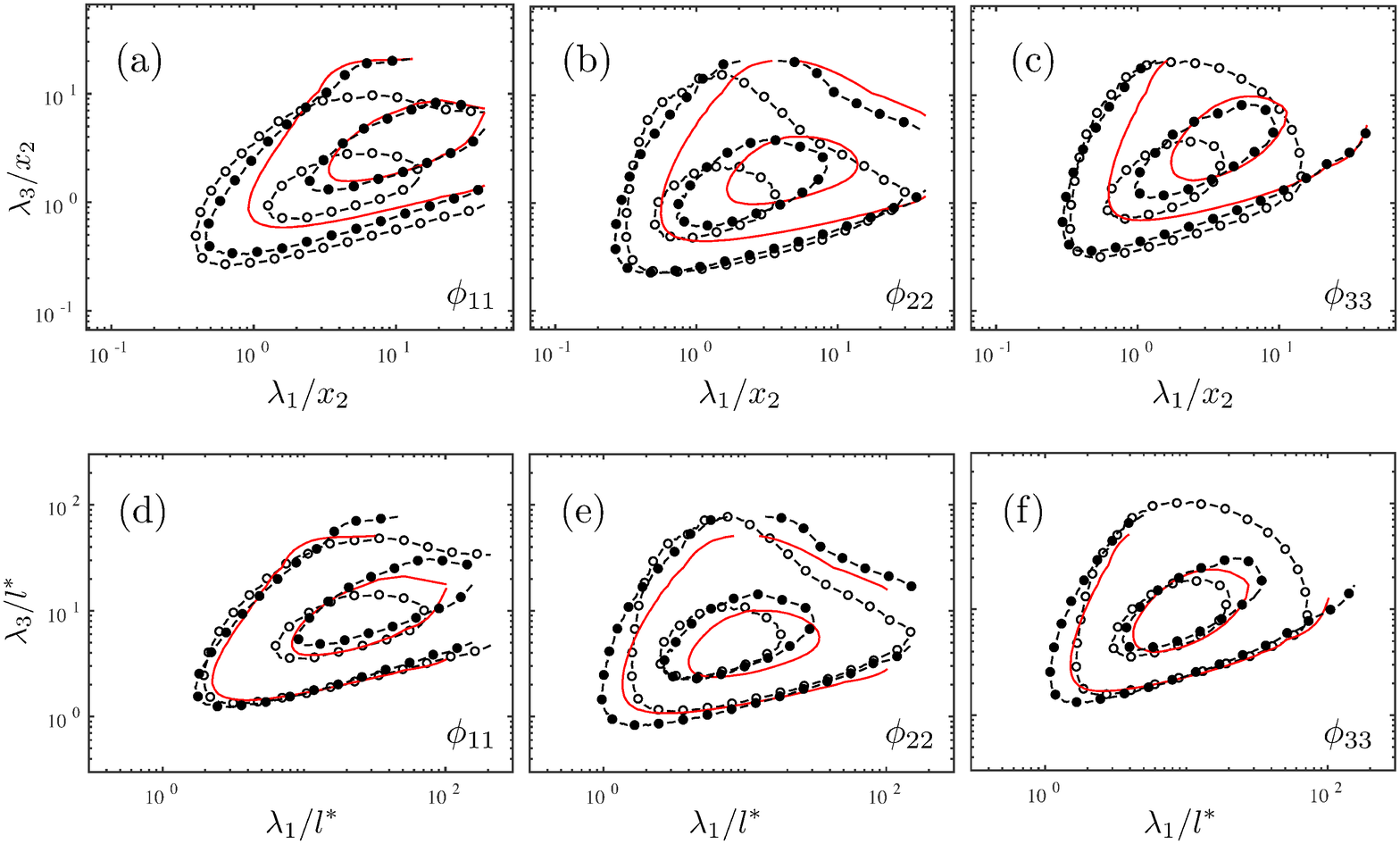}
 \end{center}
\caption{ Premultiplied streamwise $\phi_{11}$ (a,d), wall-normal
  $\phi_{22}$ (b,e), and spanwise $\phi_{33}$ (c,f) velocity
  two-dimensional spectra at $x_2=0.10h$ for NS550
  (\textcolor{red}{\solid}), NS550-s1 ($\circ$), and NS550-s2
  (\textbullet). The wavelengths are scaled by $x_2$ in the top panels
  and by $l^*$ in the bottom panels. Contours are 0.1 and 0.6 of the
  maximum.
\label{fig:Umean_fixed_spectra}}
\end{figure}

Scaling (\ref{eq:l_star}) is investigated in cases NS550, NS550-s1, and
NS550-s2 with mean velocity profiles shown in figure
\ref{fig:Umean_fixed}(a). Figure \ref{fig:Umean_fixed}(b) contains the
tangential Reynolds stress artificially generated to sustain $\langle
u_1 \rangle$.
%
%
The relevant length scale of the momentum-carrying eddies is examined
in figure \ref{fig:Umean_fixed_spectra} by comparing the
premultiplied, two-dimensional velocity spectra at $x_2=0.10h$ as a
function of the streamwise and spanwise wavelengths ($\lambda_1$ and
$\lambda_3$) scaled by the distance to the wall (top panels) and $l^*$
(bottom panels). The spectra display a noticeable mismatch when the
wavelengths are scaled by $x_2$, whereas the collapse is appreciably
improved when $\lambda_1$ and $\lambda_3$ are normalised by $l^*$,
especially for the most intense spectral cores.  Therefore, $l^*$
stands as a more faithful characterisation of the eddy sizes compared
to the distance to the wall. 
%

\begin{figure}
 \vspace{0.1cm}
 \begin{center}  
   \includegraphics[width=0.95\textwidth]{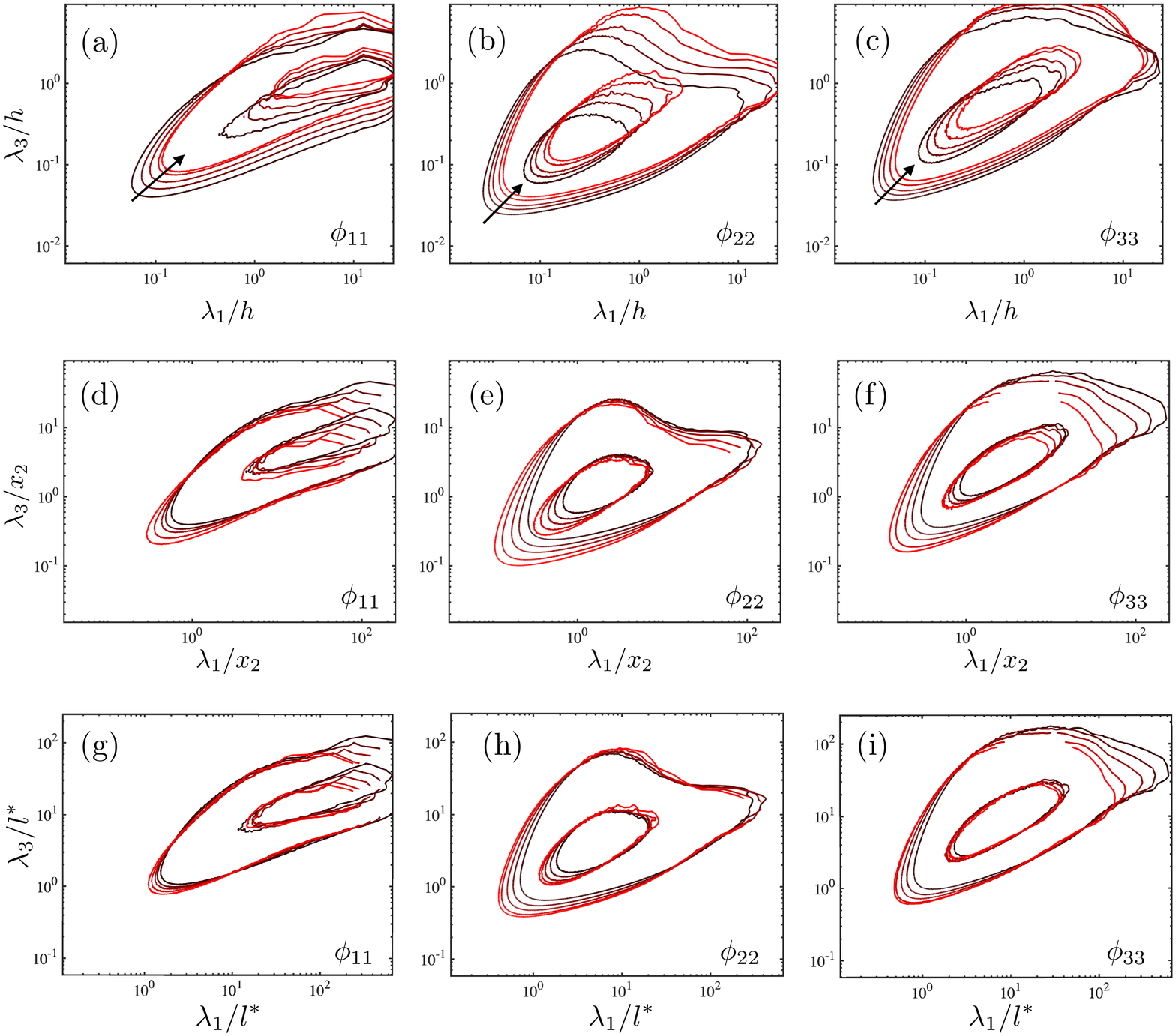}
 \end{center}
\caption{ Premultiplied streamwise $\phi_{11}$ (a,d,g), wall-normal
  $\phi_{22}$ (b,e,h), and spanwise $\phi_{33}$ (c,f,i)
  two-dimensional velocity spectra at $x_2/h=0.10,0.15,0.20,0.30$ and
  $0.40$ (from black to red) for NS2000. The wavelengths are scaled by
  $h$ (a,b,c), $x_2$ (d,e,f), and $l^*$ (g,h,i). Contours are 0.1 and
  0.6 of the maximum. The arrows in panels (a), (b), and (c) indicate
  increasing distance from the wall.
\label{fig:spectra_NS2000}}
\end{figure}

We further examine the performance of $l^*$ at different wall-normal
heights using the traditional channel flow NS2000. Figure
\ref{fig:spectra_NS2000} shows the premultiplied velocity spectra for
multiple wall-normal distances ranging from $x_2=0.1h$ to
$x_2=0.4h$. We test three different length scales to normalise the
streamwise and spanwise wavelengths, namely, $h$ (figures
\ref{fig:spectra_NS2000} a,b,c), $x_2$ (figures
\ref{fig:spectra_NS2000} d,e,f), and $l^*$ (figures
\ref{fig:spectra_NS2000} g,h,i). Again, the best collapse is attained
for $l^*$, although $x_2$ also provides a quantitative improvement
with respect to $h$.
%

The normalisation by $l^*$ is still far from perfect, although this
may be expected considering that the new scaling is only applicable to
the active motions responsible for carrying the tangential momentum
flux. For that reason, both the lower spectral contours and the
large-scale tails of $\phi_{11}$ and $\phi_{33}$ are not required to
scale with $l^*$, the former due to contamination from small eddies
decoupled from the mean shear, and the latter due to their lack of
tangential stress despite their non-zero energy content.

In summary, we have shown that $u^*$ (or $u^\star$) and $l^*$ are
tenable candidates to represent the characteristic scales of the
momentum-carrying eddies in wall-bounded turbulence. Although we have
focused on channel flows, we expect $u^*$ and $l^*$ to perform
similarly in turbulent boundary layers. The proposed scales are still
consistent with the classic scaling provided by $u_\tau$ and $x_2$ for
canonical flows, and can be considered as an extension for more
general shear-dominated turbulence.

\section{Turbulent channel with Robin boundary conditions}\label{sec:results:slip}

In this section, we analyse the significance of the distance to the
wall for the outer flow by using turbulent channel flows where the
no-slip wall is replaced by Robin boundary conditions. The new set-up
allows for instantaneous velocities at the boundaries and, in
particular, for wall-normal transpiration. No transpiration is
considered to be the most distinctive feature of walls, and it is
commonly understood as the mean by which the log-layer motions
``feel'' the distance to the wall. Hence, the non-zero $u_2$ at
$x_2=0$ (and $x_2=2h$) introduced by the Robin boundary condition is
intended to assess the role of impermeable walls as organising agents
of wall-attached eddies.

\subsection{Numerical experiments of channels with Robin boundary conditions}\label{subsec:numerical:slip}

We perform a set of DNS of turbulent channel flows using the same
numerical scheme and computational domain from \S
\ref{subsec:numerical:code}. The no-slip wall is replaced by a Robin
boundary condition of the form
\begin{equation}
u_i|_w = l \left.\frac{\partial u_i}{\partial
n}\right|_w,\quad i=1,2,3,
\label{eq:slip_bc}
\end{equation}
where the subscript $w$ refers to quantities evaluated at the wall,
and $n$ is the wall-normal (or boundary-normal) direction. We define
$l$ to be the slip length that, in general, may be a function of the
spatial wall-parallel coordinates and time. The choice of $l$ must
comply with the symmetries of the flow and, particularly for a channel
flow configuration, (\ref{eq:slip_bc}) should satisfy
\begin{equation}
\langle u_i|_w\rangle = \left\langle\left.  l\frac{\partial
  u_i}{\partial x_2}\right|_w\right\rangle = 0,\quad i=2,3.
\label{eq:avg_bc}
\end{equation}
In the present study, we consider a constant value for $l$. This is
consistent with (\ref{eq:avg_bc}) because $\langle u_i |_w \rangle =
0$ and $\left.\left\langle{\partial u_i}/{\partial x_2} \right|_w
\right\rangle = 0$ for $i=2,3$. The cases simulated are summarised in
table \ref{tab:cases:slip}.  All the simulations were run for at least
10 eddy turnover times after transients.  Throughout the text, we
occasionally refer to cases with the Robin boundary condition as
Robin-bounded and those with the no-slip condition as
wall-bounded. Robin boundary conditions have been previously employed
in the wall-parallel directions to model the flow over hydrophobic
surfaces
\citep{Min2004,Martell2009,Park2013,Jelly2014,Seo2015,Seo2016}, but
note that in the present study the boundary condition is also applied
to the wall-normal velocity, which is seldom done in the
literature. The slip lengths used here are also larger than the
typical values in hydrophobic works.
%
\begin{table}
\begin{center}
\setlength{\tabcolsep}{12pt}
\begin{tabular}{l c c c c c c c c}
Case       &  $\Rey_\tau$ & $\Delta x_1^+$  & $\Delta x^+_{2,\mathrm{min}/\mathrm{max}}$  & $\Delta x_3^+$ & $l/h$ & Driven by \\
\hline
\hline
R550        & 546          & 6.7   & 0.2/9.9  & 3.3   & 0.10 &  constant $\langle\frac{\mathrm{d}p}{\mathrm{d}x_1}\rangle$ \\
R550-l1     & 546          & 6.7   & 0.2/9.9  & 3.3   & 0.25 &  constant $\langle\frac{\mathrm{d}p}{\mathrm{d}x_1}\rangle$ \\
R550-l2     & 546          & 6.7   & 0.2/9.9  & 3.3   & 0.50 &  constant $\langle\frac{\mathrm{d}p}{\mathrm{d}x_1}\rangle$ \\
R950        & 934          & 5.7   & 0.5/10.1 & 2.8   & 0.10 &  constant $\langle\frac{\mathrm{d}p}{\mathrm{d}x_1}\rangle$ \\
R2000       & 2003         & 6.1   & 0.7/15.0 & 3.1   & 0.10 &  constant $\langle\frac{\mathrm{d}p}{\mathrm{d}x_1}\rangle$ \\
\hline
\end{tabular}
\end{center}
\caption{Tabulated list of cases for \S \ref{sec:results:slip}. The
  numerical experiments are labelled following the convention
  R[$\Rey_\tau$]-[specific case], where R denotes channel with Robin
  boundary condition.  $\Delta x_1$, $\Delta x_2$ and $\Delta x_3$ are
  the streamwise, wall-normal, and spanwise grid resolutions,
  respectively. The slip length used in the Robin boundary condition
  is $l$. The last column shows the method employed to drive the
  channel flow. See text for details.}
\label{tab:cases:slip}
\end{table}

\begin{figure}
\begin{center}
 \vspace{0.1cm}
 \psfrag{X}{$x_2^+$}\psfrag{Y}{$-\langle u_1^+ u_2^+ \rangle$}
 \subfloat[]{\includegraphics[width=0.45\textwidth]{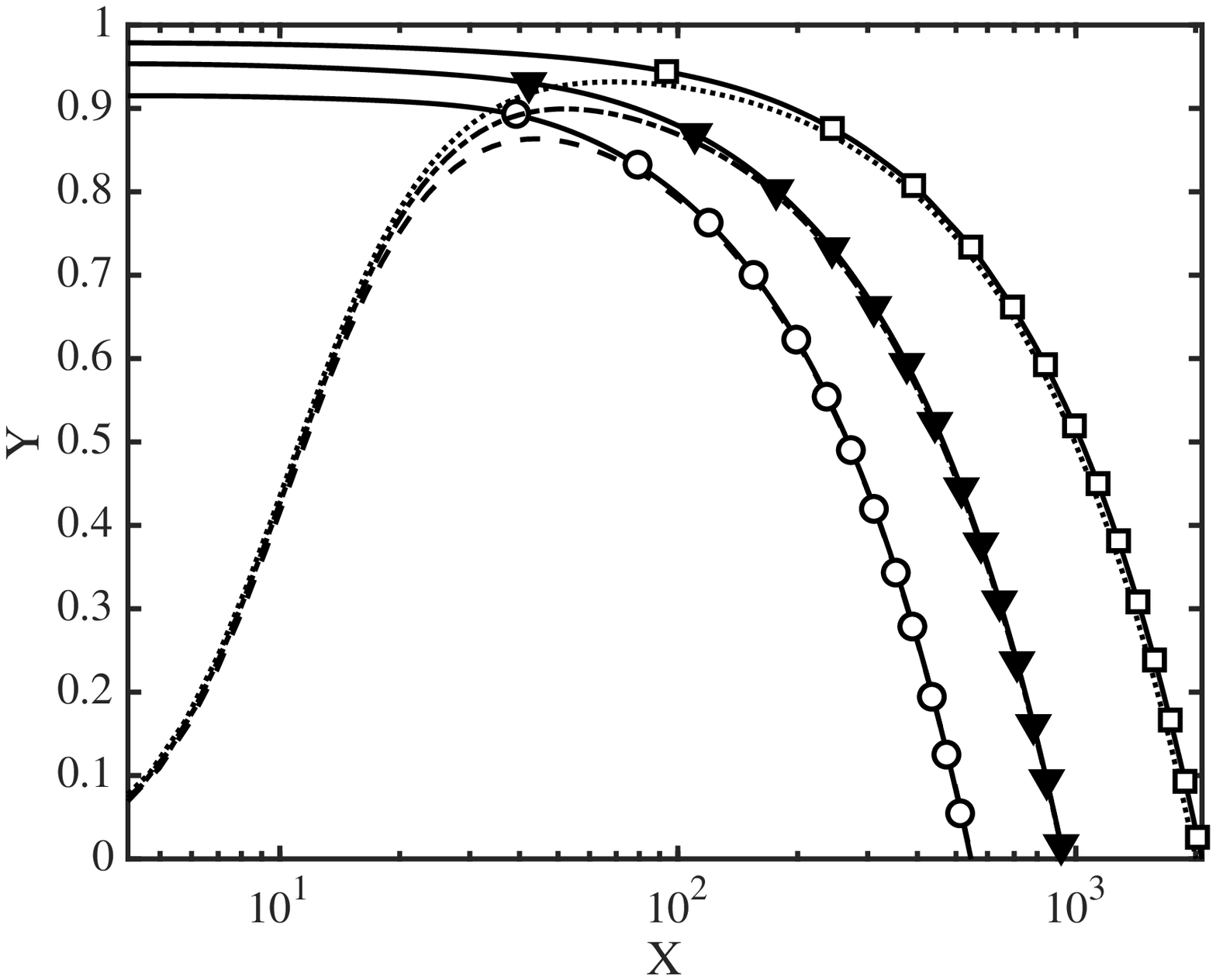}}
 \hspace{0.1cm}
 \psfrag{X}{$\lambda_1/h$}\psfrag{Y}{$\lambda_3/h$}
 \subfloat[]{\includegraphics[width=0.47\textwidth]{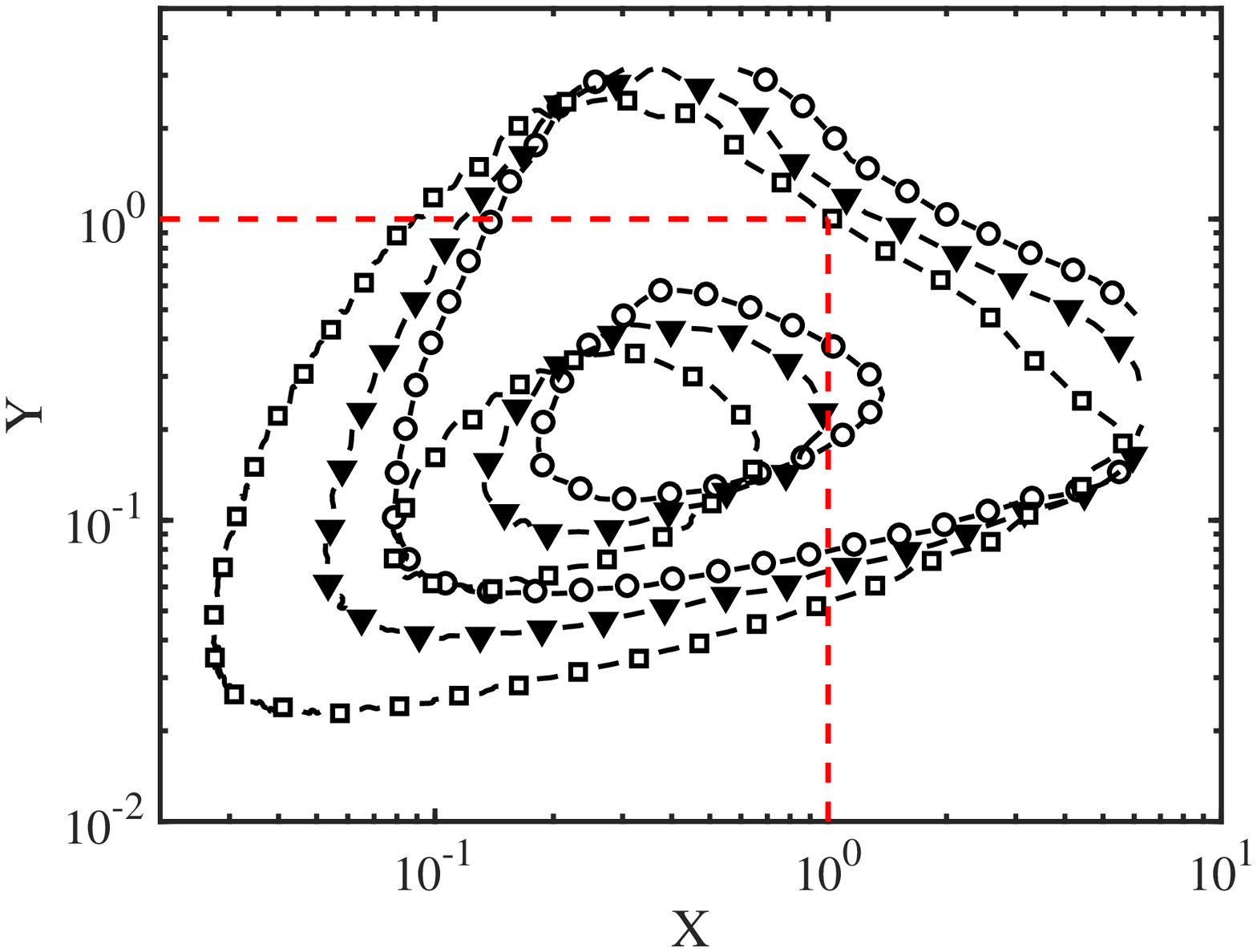}}
\end{center}
\caption{ (a) Mean tangential Reynolds stress as a function of the
  wall-normal coordinate for R550 ($\circ$), R950
  ($\blacktriangledown$), R2000 ($\square$), NS550 (\dashed), NS950
  (\dchndot), and NS2000 (\dotted). (b) Premultiplied wall-normal
  two-dimensional velocity spectra for Robin-bounded channels at
  $x_2/h=0$.  Contours are 0.1 and 0.6 of the maximum. The red dashed
  lines are $\lambda_1/h=1$ and $\lambda_3/h=1$. Symbols are as in
  (a).
\label{fig:mean_uv_spectra}}
\end{figure}

The motivation of using (\ref{eq:slip_bc}) is to provide a boundary
for the flow that deviates from the behaviour of a regular
wall. Indeed, for large values of $l$, (\ref{eq:slip_bc}) constitutes
a significant modification of the classic no-slip boundary condition
by suppressing the formation of near-wall viscous layers
\citep{Lozano2016_Brief}.
The mean tangential Reynolds stress is shown in figure
\ref{fig:mean_uv_spectra}(a) for cases R550, R950, and R2000 with slip
length $l=0.10h$. For the three Reynolds numbers under consideration,
$-\langle u_1 u_2 \rangle$ captures more than 90\% of the total
stress, and this was the criteria used to select $l=0.10h$ as the
reference slip length. As the Reynolds number increases, so does the
contribution of $-\langle u_1 u_2 \rangle$ close to the wall at the
expense of reducing the formation of near-wall viscous layers that
appear prior to the proximity of the wall. Far from the boundaries,
the tangential momentum flux is linear for the same reason as in
wall-bounded channels, i.e., to balance the constant mean pressure
gradient driving the flow. Cases R550-l1 and R550-l2 are for $l=0.25h$
and $l=0.50h$, respectively, and are intended to test the effect of
increasing slip lengths.
%

Another two important properties of the Robin boundary condition are
that it allows for transpiration at all flow scales and, it does not
encode any specific information regarding the linear wall-normal
scaling of the log-layer eddies.  The spectral density of the
wall-normal velocity, $\phi_{22}$, evaluated at $x_2/h = 0$ for
Robin-bounded cases is shown in figure \ref{fig:mean_uv_spectra}(b) as
a function of the streamwise and spanwise wavelengths, $\lambda_1$ and
$\lambda_3$, respectively.  The spectra are non-zero at the boundary
with a non-negligible contribution from wavelengths up to $\lambda_1$
and $\lambda_3$ of $\mathcal{O}(h)$. Moreover, the spectral energies
obtained by integrating $\phi_{22}$ at $x_2/h=0$ are approximately
$u_\tau^2$, that is of the same order as the values in the bulk
flow. This implies that the Robin boundary should alter the behaviour
of eddies with sizes up to $\mathcal{O}(h)$ if they are controlled by
the distance to the wall as commonly hypothesised
\citep{Townsend1976}.

One could still argue that $\langle u_2|_w \rangle$ is zero,
analogously to the scenario encountered for impermeable
walls. However, note that this is also the case for $\langle u_2
\rangle$ at each wall-normal location for a no-slip channel, and that
(\ref{eq:avg_bc}), and hence $\langle u_2|_w \rangle = 0$, is just the
immediate consequence of the symmetries of the channel flow
configuration rather than a result of the impermeability constraint.
The only reminiscence of the wall in the Robin boundary condition
comes from the fact that, by construction of (\ref{eq:slip_bc}), the
velocities and their corresponding wall-normal derivatives are
expected to show similar lengths scales in the vicinity of the wall
akin to the production and dissipation in the buffer region of a
smooth wall.

\begin{figure}
\begin{center}
\vspace{0.1cm}
\includegraphics[width=0.95\textwidth]{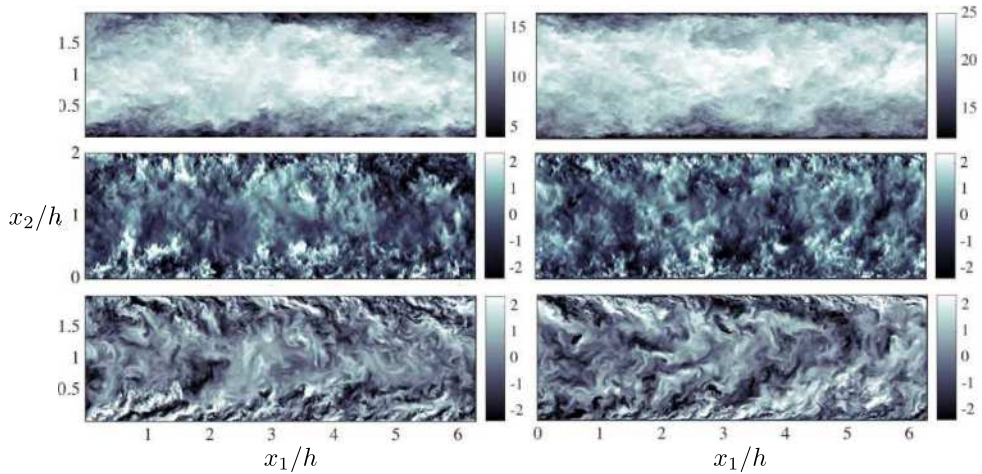}
\end{center}
\caption{Instantaneous $x_1$--$x_2$ planes of the streamwise (top),
  wall-normal (centre), and spanwise (bottom) velocities for turbulent
  channels at $Re_\tau=2000$. Left panels are for R2000, and right
  panels are for NS2000. The colour bars show velocity in wall
  units.\label{fig:snapshots_slip}}
\end{figure}

Instantaneous flow fields for the three velocity components at
$Re_\tau=2000$ for NS2000 and R2000 are compared in figure
\ref{fig:snapshots_slip}. The similarities between both visualisations
are striking, and their resemblance is confirmed by the quantitative
analysis of the flow statistics presented in the following sections.
Note that the Robin boundary condition is introduced in the present
work just as a mean to confine the flow within a boundary which
differs from a no-slip wall, especially regarding transpiration. In
this sense, the Robin boundary condition is just a tool and it is not
intended to model any particular flow phenomena.  Moreover, the focus
of this paper lies on the outer region and we are less interested in
the structure of the flow at the boundary. Nonetheless, a further
characterisation of the flow at $x_2/h = 0$ is provided in Appendix
\ref{sec:appendixA} for completeness.
%

\subsection{One-point statistics and spectra}\label{subsec:results:stats}

\begin{figure}
\begin{center}
 \vspace{0.1cm}
 \psfrag{X}[br]{$x_2/h$}\psfrag{Y}{$\langle u_1^+ \rangle$}
 \subfloat[]{ \includegraphics[width=0.47\textwidth]{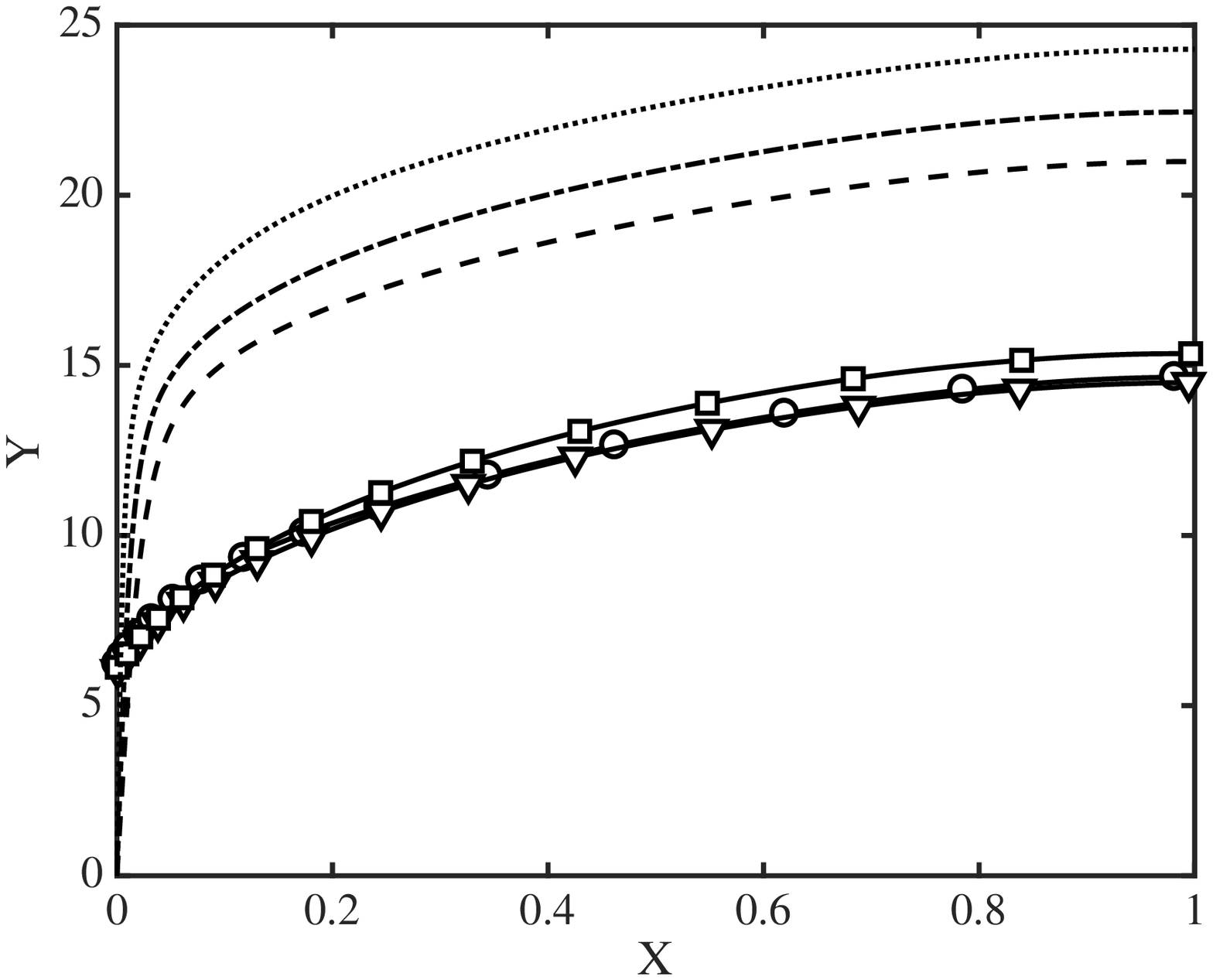} }
 \hspace{0.2cm}
 \psfrag{X}[br]{$x_2^+$}\psfrag{Y}{$\langle u_1^+ \rangle - u_1^{\mathrm{shift}}$}
 \subfloat[]{ \includegraphics[width=0.47\textwidth]{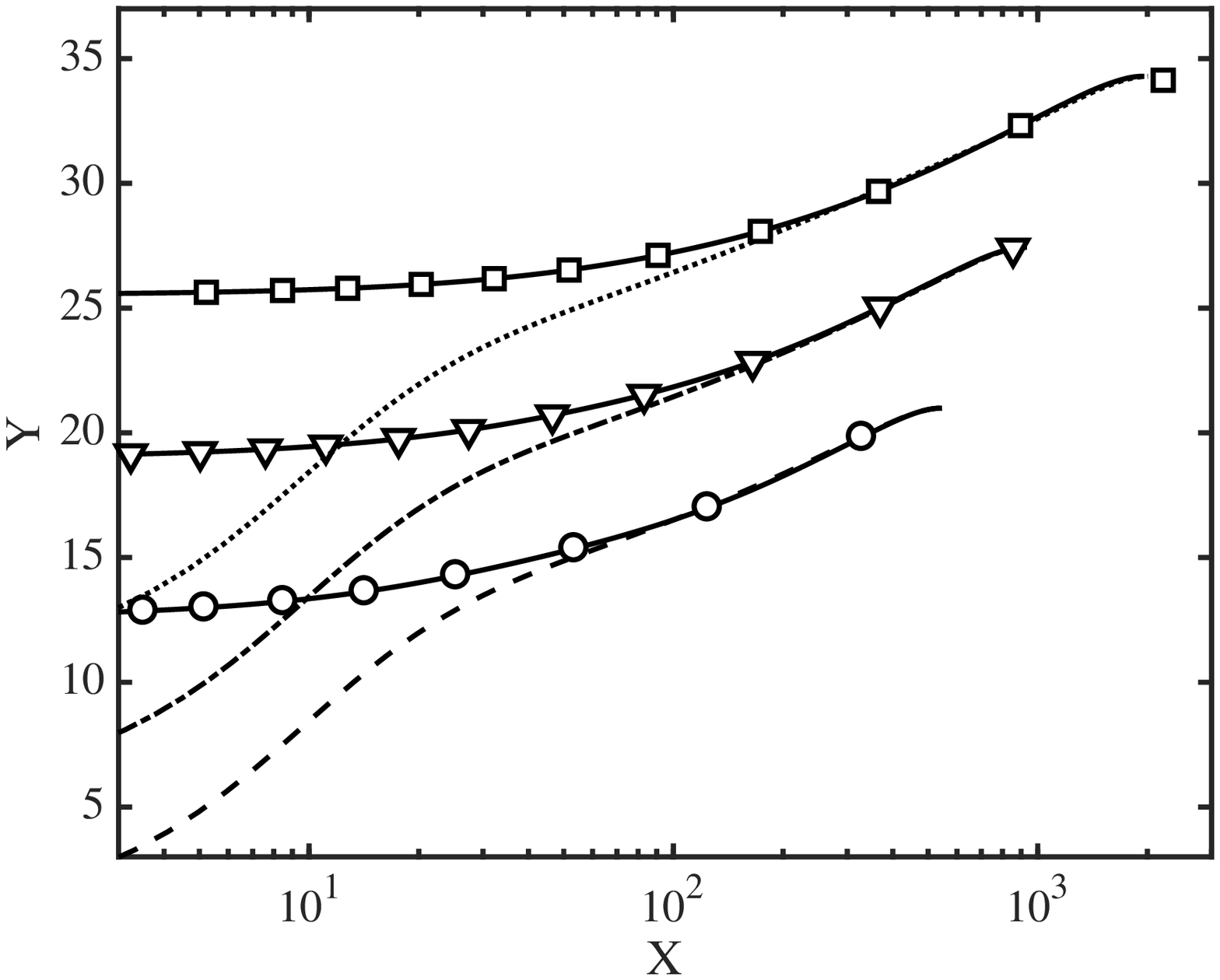} }
 \end{center}
\caption{ Mean streamwise velocity profiles as a function of $x_2$ in
  (a) linear scale and (b) logarithmic scale for R550 ($\circ$), R950
  ($\triangledown$), R2000 ($\square$), NS550 (\dashed), NS950
  (\dchndot), and NS2000 (\dotted). In panel (b), the velocity
  profiles of Robin-bounded cases are vertically shifted such that
  centreline velocity coincide with their corresponding wall-bounded
  case.  For clarity, profiles at $Re_\tau \approx 950$ and $Re_\tau
  \approx 2000$ are additionally shifted by 5 and 10 plus units,
  respectively.\label{fig:mean}}
\end{figure}

The mean streamwise velocities for the Robin-bounded and wall-bounded
channel flows are compared in figure \ref{fig:mean}. In figure
\ref{fig:mean}(b), the mean profiles for Robin-bounded cases are
vertically displaced to match the centreline velocity of the
corresponding no-slip case. A first observation is that the shape of
$\langle u_1\rangle$ remains roughly identical for $x_2 \gtrsim
0.10h$, and the Robin boundary condition is mainly responsible for a
reduction of the total mass flux.  The shifts required to match the
Robin-bounded cases to their no-slip counterpart were positive and
equal to $6.4$, $8.0$, and $8.9$ plus units for R550, R950, and R2000,
respectively. Nonetheless, we will not emphasise these values as they
can be trivially changed by either adding a constant uniform velocity
to the right-hand side of the Robin boundary condition
(\ref{eq:slip_bc}), or by a Galilean transformation of the velocity
field.

The observations from figure \ref{fig:mean} can be connected to the
law of the wall
\citep{Prandtl1925,Karman1930,Millikan1938,Townsend1976},
\begin{equation}
\langle u_1^+ \rangle = \frac{1}{\kappa}
\log\left( x_2^+ \right) + B + \Pi,
\label{eq:log_law}
\end{equation}
where $\kappa$ and $B$ are the von K\'arm\'an and intercept constants
for no-slip walls, respectively, and $\Pi$ is an additional velocity
displacement. The friction velocity for Robin-bounded cases is $\utau
= \sqrt{\nu \partial \langle u_1 \rangle/\partial x_2 - \langle u_1
  u_2\rangle}|_{x_2=0}$. Hence, the Robin boundary condition
introduces a non-zero tangential Reynolds stress at the wall, which
acts as an effective drag with a major impact on $\Pi$
\citep[see][]{Nikuradse1933,Jimenez2004}. On the hand, we can
hypothesise that the $x_2$-dependent component, $\sim 1/\kappa
\log(x_2)$, is mainly controlled by the momentum flux for $x_2>0$.
%

\begin{figure}
\begin{center}
 \vspace{0.1cm}
 \psfrag{X}[cr]{$x_2/h$}\psfrag{Y}{$\langle {u_1'}^{2+} \rangle^{1/2}$}
 \subfloat[]{  \includegraphics[width=0.48\textwidth]{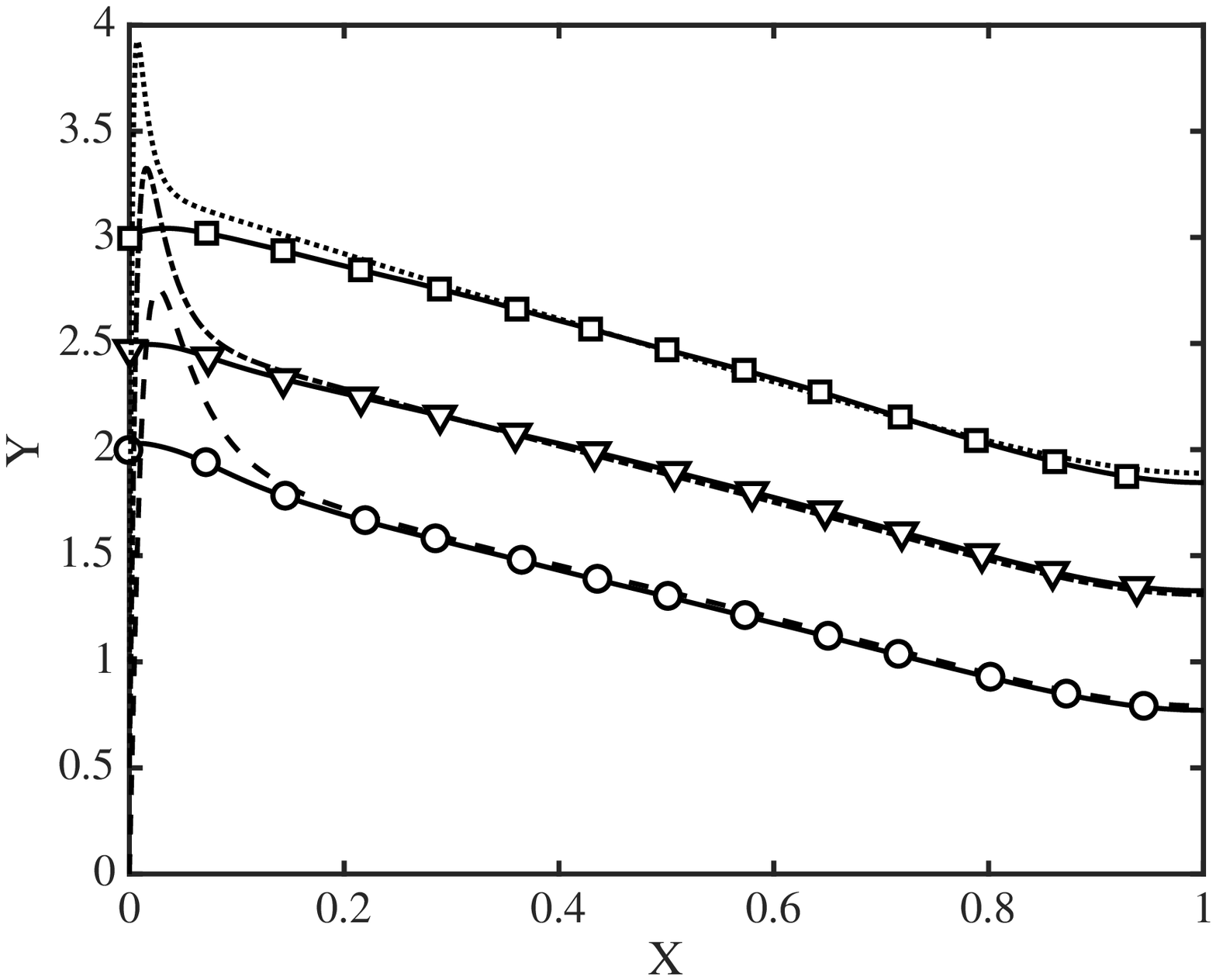} }
 \hspace{0.01cm}
 \psfrag{X}[cr]{$x_2/h$}\psfrag{Y}{$\langle {u_2'}^{2+} \rangle^{1/2}$}
 \subfloat[]{ \includegraphics[width=0.48\textwidth]{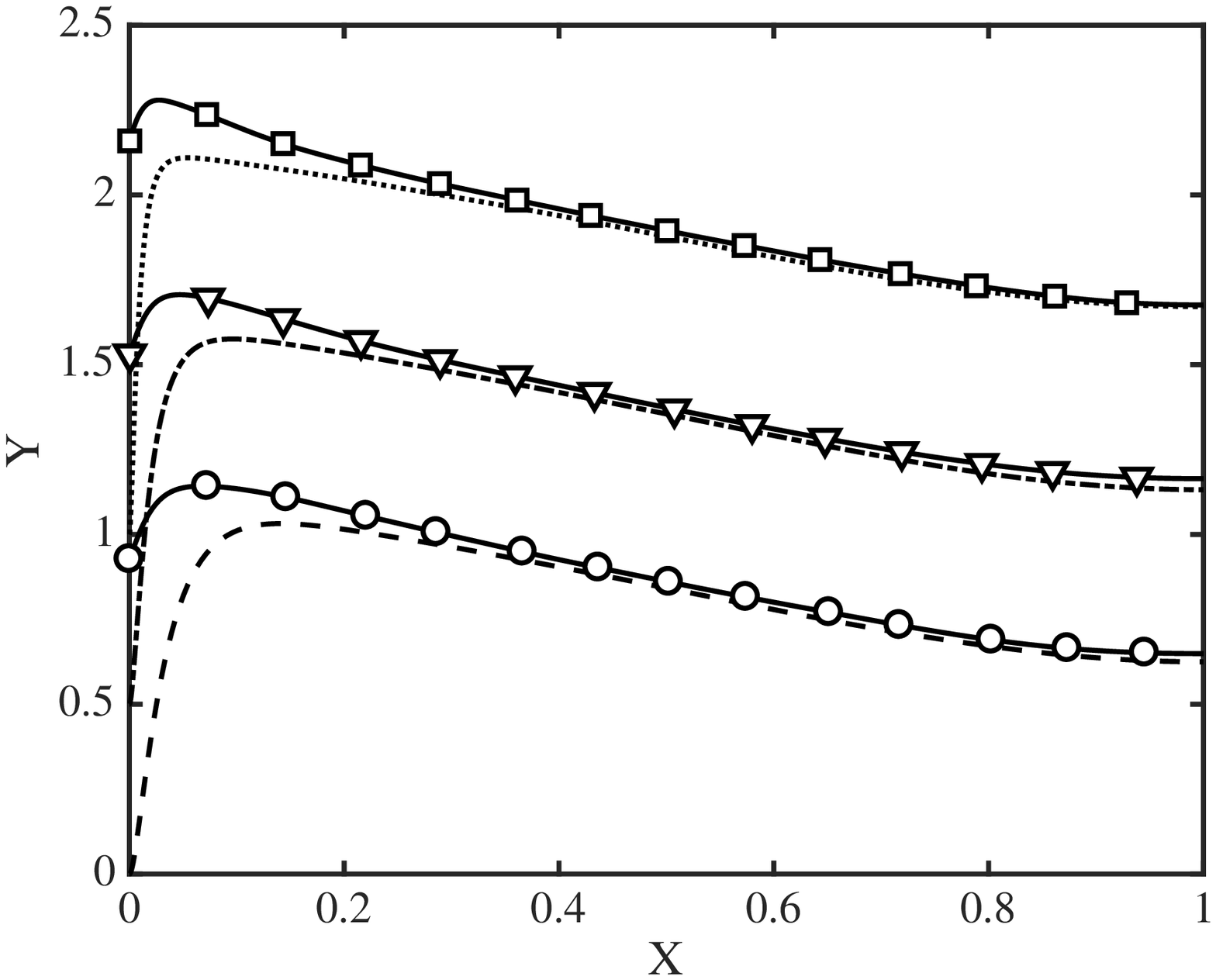} }
 \end{center}
%
%
 \begin{center}  
   \psfrag{X}[cr]{$x_2/h$}\psfrag{Y}{$\langle {u_3'}^{2+} \rangle^{1/2}$}
   \subfloat[]{ \includegraphics[width=0.48\textwidth]{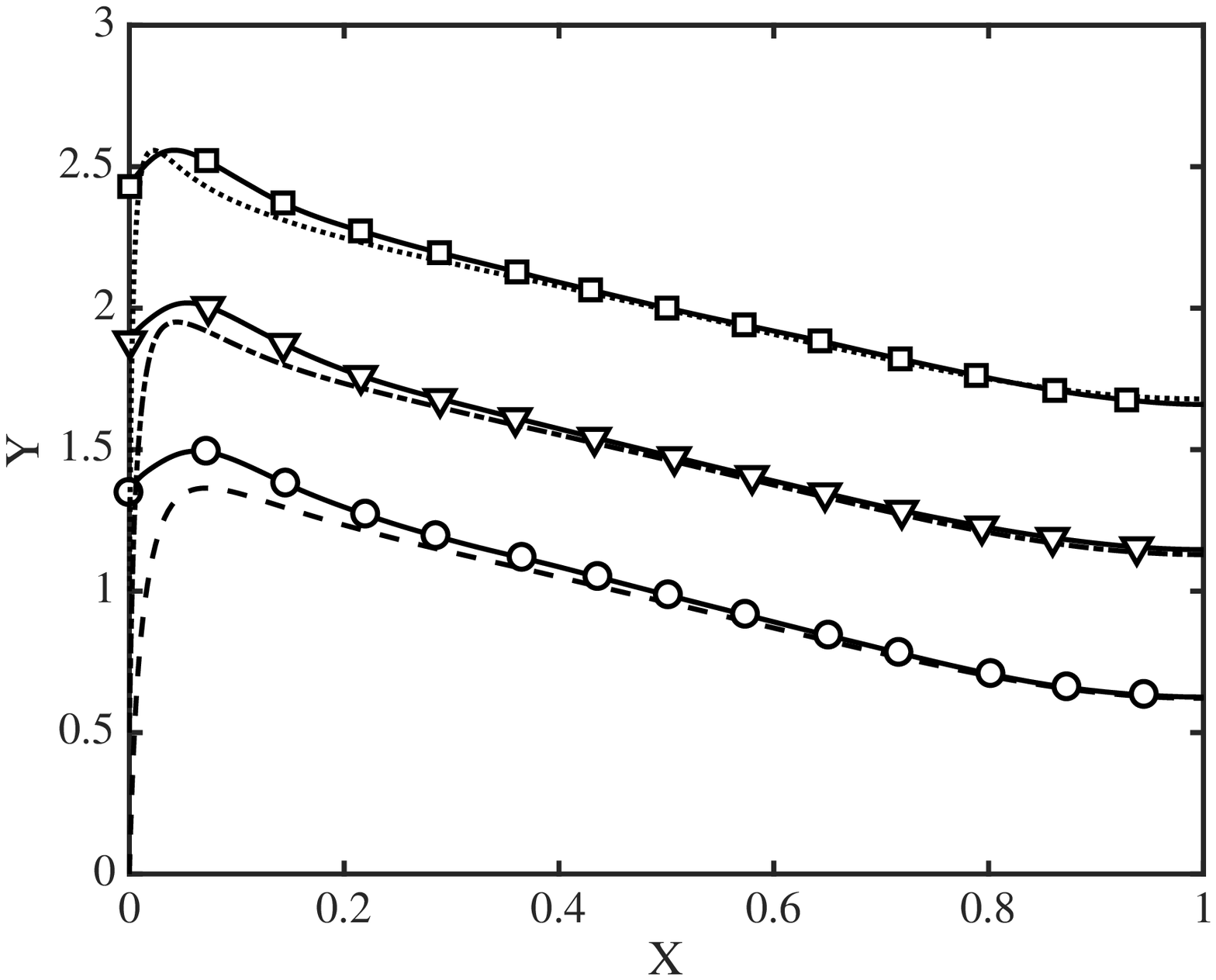} }
   \hspace{0.01cm}
   \psfrag{X}[cr]{$x_2/h$}\psfrag{Y}{$\langle {p'}^{2+} \rangle^{1/2}$}
   \subfloat[]{ \includegraphics[width=0.48\textwidth]{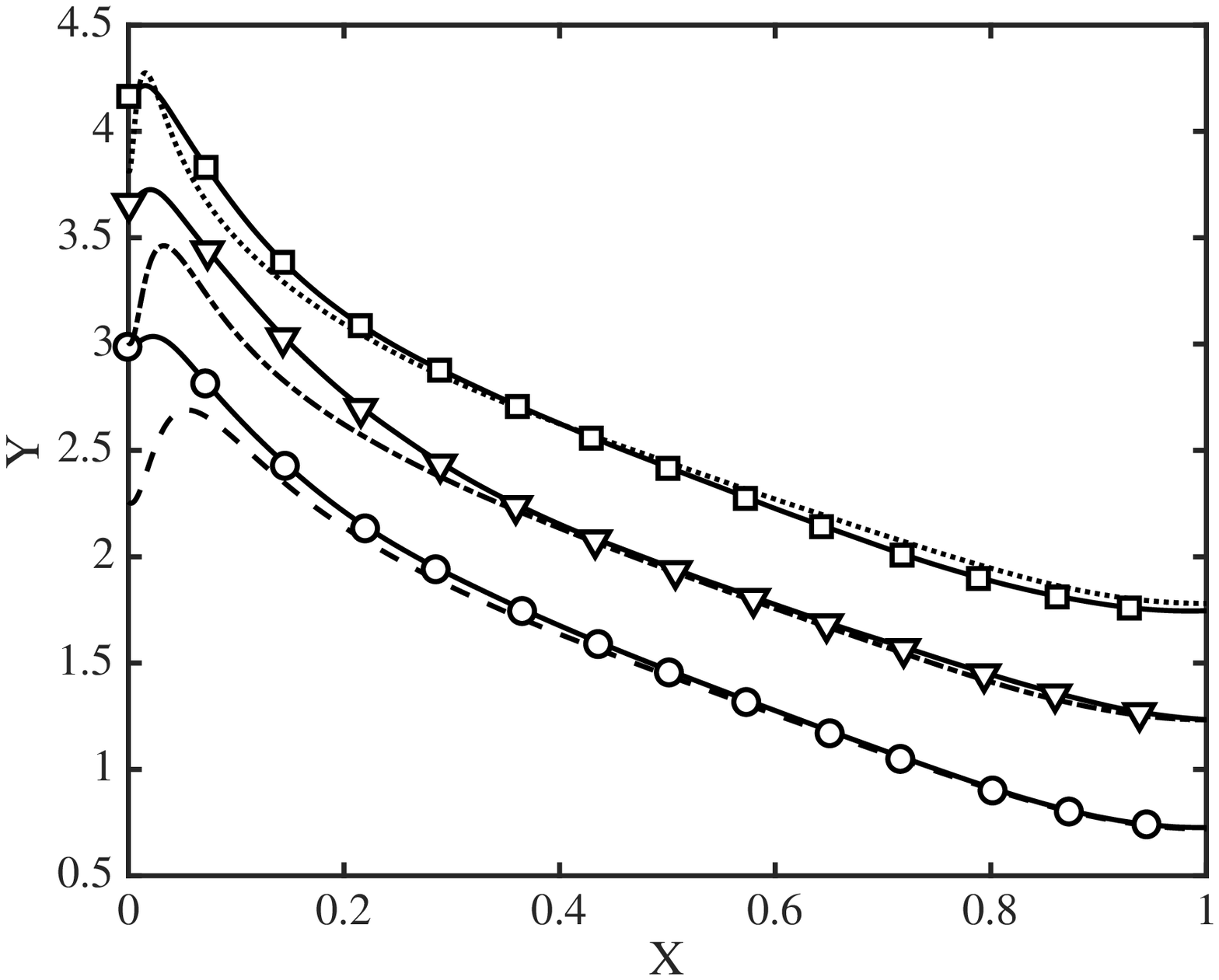} }
 \end{center}
\caption{ (a) Streamwise, (b) wall-normal, and (c) spanwise
  root-mean-squared velocity fluctuations, and (d) root-mean-squared
  pressure fluctuations for R550 ($\circ$), R950 ($\triangledown$),
  R2000 ($\square$), NS550 (\dashed), NS950 (\dchndot), and NS2000
  (\dotted). In all panels, the profiles for cases at $Re_\tau \approx
  950$ and $Re_\tau \approx 2000$ are vertically shifted by $0.5$ and
  $1$ plus units, respectively, for clarity.\label{fig:rms}}
\end{figure}

The r.m.s. velocity fluctuations for the Robin-bounded and
wall-bounded channels are shown in figure \ref{fig:rms}(a--c).  The
lack of a near-wall peak at $x_2^+\approx 15$ for the Robin-bounded
streamwise velocity fluctuations is a consequence of the interruption
of the classic near-wall cycle \citep{Jimenez1991,Jimenez1999}, which
is also confirmed by visual inspection of the instantaneous near-wall
velocities (see Appendix \ref{sec:appendixA}).  The most remarkable
observation from figure \ref{fig:rms} is that the Robin-bounded
fluctuating velocities match quantitatively their wall-bounded
counterparts for $x_2 \gtrsim 0.10h$ despite the lack of impermeable
walls. The presence of a significant non-zero $u_2$ in the Robin-bounded
cases is evidenced by the r.m.s. of $u_2$ at $x_2/h=0$ whose values are
comparable to the r.m.s. in the bulk flow. The result is again an
indication that the total contribution of the different eddies to the
turbulence intensities is insensitive to the presence of impermeable
walls.  The pressure fluctuations are also accurately reproduced in
figure \ref{fig:rms}(d) for the Robin-bounded cases.  The results may
appear surprising due to the global nature of the pressure, which is
more prone to being affected by the changes in the boundary
condition. However, the most important contribution of the pressure is
to guarantee local continuity among eddies, and the satisfactory
collapse of the r.m.s.  velocity fluctuations seen in figures
\ref{fig:rms}(a--c) also imply equally fair results for the pressure
fluctuations.  This is in accordance with \cite{Sillero2014}, who
concluded from the pressure correlations in boundary layers that $p$
is dominated by localised regions of strongly coupled small-scale
structures.
%

\begin{figure}
\begin{center}
 \vspace{0.1cm}
 \includegraphics[width=1\textwidth]{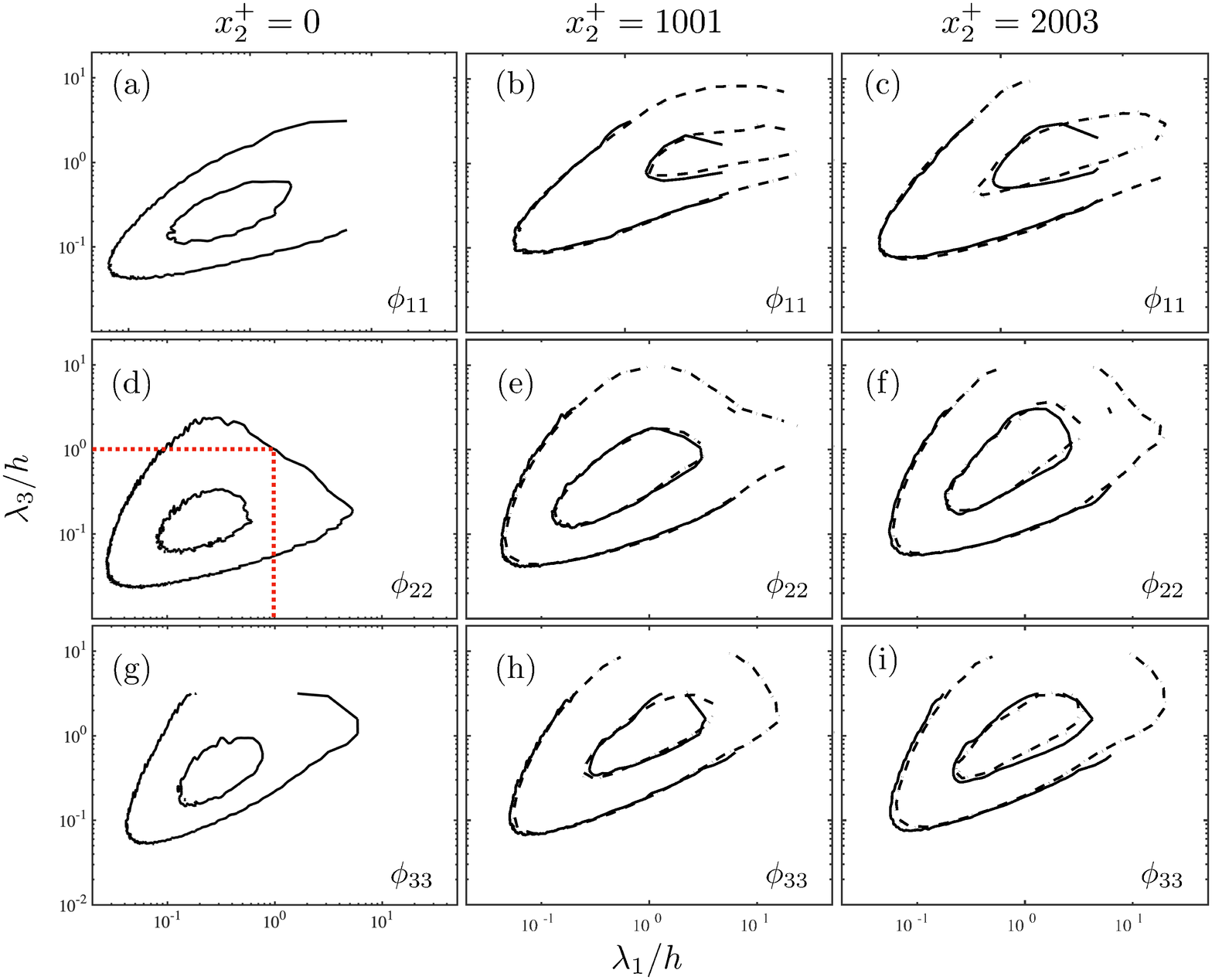}
\end{center}
\caption{Wall-parallel premultiplied streamwise (a,b,c), wall-normal
  (d,e,f), and spanwise (g,h,i) velocity spectra for R2000 (\solid) and
  NS2000 (\dashed). The wall-normal locations are $x_2^+=0$ (a,d,g),
  $x_2^+=1001$ (b,e,h), and $x_2^+=2003$ (c,f,i). Contours are 0.1 and
  0.6 of the maximum. The red dotted lines in (d) are $\lambda_1/h =
  1$ and $\lambda_3/h = 1$.
\label{fig:spectra}}
\end{figure}

The spectral densities of the three velocity fluctuations,
$\phi_{11}$, $\phi_{22}$, and $\phi_{33}$, are shown in figure
\ref{fig:spectra} as a function of the streamwise and spanwise
wavelengths. Several wall-normal heights are considered for R2000 and
NS2000.  The spectra for wall-bounded cases is zero at
$x_2/h=0$. Conversely, $\phi_{22}$ is non-zero at the boundary for the
Robin-bounded case, and peaks at $\lambda_1\approx 0.3h$ and
$\lambda_3\approx 0.10h$, with a non-negligible contribution from
wavelengths up to $\lambda_1$ and $\lambda_3$ of $\mathcal{O}(h)$. We
can then estimate the flow scales that are expected to be affected by
the transpiration of the boundary by assuming that the stress-carrying
eddies follow $\lambda_1 \approx 1.5 x_2$ and $\lambda_3 \approx x_2 $
(as shown below in \S \ref{subsec:results:attached}). If the
boundary is perceived as permeable for scales up to $\mathcal{O}(h)$,
then attached motions below $x_2 \approx 0.7h$ should adjust
accordingly to accommodate transpiration effects, especially if the
wall is their primary organising agent.  However, inspection of the
spectra above $x_2 \approx 0.10h$ shows that the agreement between
wall-bounded and Robin-bounded channels is outstanding (figures
\ref{fig:spectra} b,c,e,f,h,i). Consequently, the distance to the
boundary (or non-existent wall) is not the relevant length scale
controlling the size of the attached eddies, in contrast with the
traditional argument by \cite{Townsend1976}. Instead, the resemblance
between wall-bounded and Robin-bounded cases presented above should be
attributed to the common momentum transfer $u_\tau^2$ characteristic
of both cases as argued in \S \ref{sec:scales}.
%

It also worth noting that the excellent agreement of the one-point
statistics and spectra occurs in spite of the absence of very
large-scale motions longer than $\sim$$6h$
\citep{Guala2006,Balakumar2007} that do not fit within the limited
computational domain of the present simulations.  The result is
consistent with \cite{Flores2010} and \cite{Lozano2014a}, and scales
larger than $\sim$$6h$ can be essentially modelled as infinitely long
(due to the streamwise periodicity of the channel) while still
interacting correctly with the smaller well-resolved eddies.

\subsection{Existence of a virtual wall vs. adaptation layer from the boundary}
\label{subsec:results:adaptation}

\begin{figure}
\begin{center}
 \vspace{0.1cm}
 \psfrag{X}{$\lambda_3/h$}\psfrag{Y}{$\lambda_1/h$}
 \subfloat[]{\includegraphics[width=0.48\textwidth]{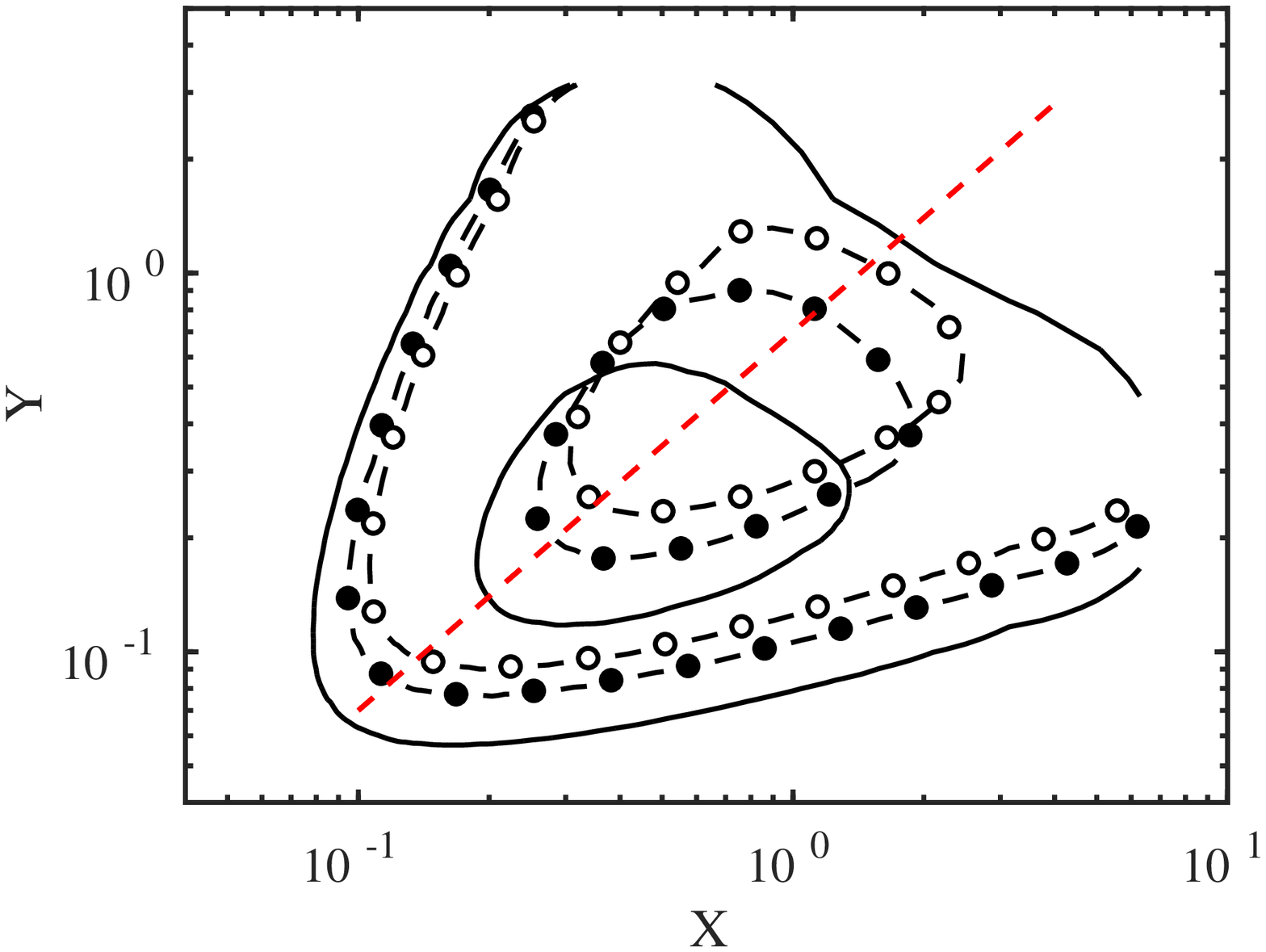} }
 \psfrag{X}{$l/h$}\psfrag{Y}{$l_a/h$}
 \subfloat[]{\includegraphics[width=0.48\textwidth]{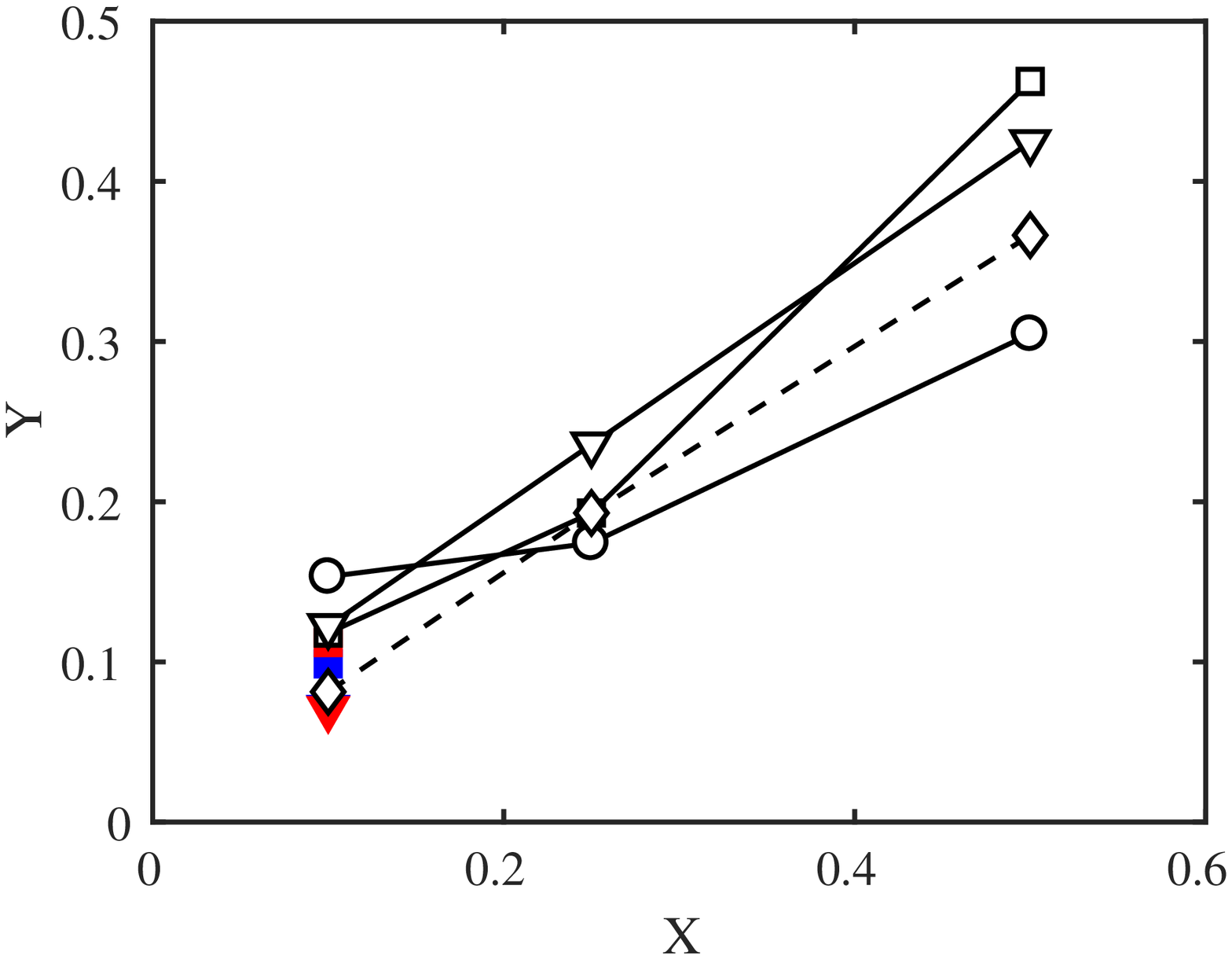} }
\end{center}
\caption{ (a) Wall-parallel premultiplied wall-normal velocity spectra
  at $x_2^+=0$ for R550 (\solid), R550-l1 ($\circ$), and R550-l2
  (\textbullet).  Contours are 0.1 and 0.6 of the maximum. The red
  dashed line is $\lambda_3 = 0.7 \lambda_1$. (b) Adaptation length
  $l_a$ for the mean velocity profile ($\Diamond$), and
  root-mean-squared streamwise ($\circ$), wall-normal ($\square$) and
  spanwise ($\triangledown$) velocity fluctuations as a function of
  the slip length $l$. Colours are black for R550, blue for R950, and
  red for R2000. \label{fig:adaptation}}
\end{figure}

It could be still argued that the flow is influenced by an effective
virtual wall whose origin is vertically displaced with respect to the
boundary. In that case, different values of the slip length $l$ would
be accompanied by changes in the origin of the virtual wall.  The
virtual-wall argument, often invoked in roughness studies
\citep{Raupach1991,Jimenez2004}, proved not to be necessary in \S
\ref{subsec:results:stats}, where it was shown that the flow recovers
above some wall-normal distance, $l_a$, without requiring a shift in
$x_2$. Nevertheless, the position of the hypothetical virtual wall,
$\delta x_2$, can be estimated by a least-square fit of the mean
velocity profile to the logarithmic law \citep{Raupach1991},
\begin{equation}
\langle  u_1^+ \rangle = \frac{1}{\kappa} \log( x_2^+ - \delta x_2^+ ) 
+ B - \Delta \langle  u_1^+ \rangle,
\label{eq:log_fit}
\end{equation}
with $\kappa = 0.392$ and $B=4.48$ \citep{Luchini2017}, within the
range $x_2 \in [0.1h,0.2h]$.  For this purpose, we will consider two
additional cases with $l=0.25h$ and $l=0.50h$ at $Re_\tau \approx 550$
labelled respectively as R550-l1 and R550-l2 in table
\ref{tab:cases:slip}. The spectral density $\phi_{22}$ for the three
different slip lengths is reported in figure \ref{fig:adaptation}(a)
at $x_2/h=0$. As $l$ increases, the spectrum moves towards higher
wavelengths along the ridge $\lambda_1 \sim \lambda_3$, and
transpiration is allowed for larger scales, which in principle should
change the location of the virtual wall. However, the virtual wall
off-set computed with (\ref{eq:log_fit}) is $\delta x_2^+ \approx
\mathcal{O}(10)$ for all the Robin-bounded cases in table
\ref{tab:cases:slip}, which is not significant enough to alter in a
meaningful manner the energy-containing eddies populating the
log-layer. Furthermore, when the $x_2$ direction was remapped into
$\tilde x_2 = x_2 + \delta x_2 (1 - x_2/h)$ \citep{Flores2006}, the
collapse of the r.m.s. velocity fluctuations worsened despite the minor
improvements in $\langle u_1 \rangle$. 
%
Consequently, the results do not comply with the existence of a the
virtual wall. Note that the same conclusion does not apply to the
viscous eddies close to the wall with sizes comparable to $\delta
x_2$, where changes in the permeability of the boundary are expected
to have a considerable impact on the near-wall flow dynamics as
typically described in flow control strategies
\citep{Choi1994,Abderrahaman2017}.

Instead of a virtual wall, we analyse the results assuming the
existence of an adaptation layer from the boundary. The vertical
distance from the boundary above which the flow recovers to the
nominal no-slip flow statistics, $l_a$, is plotted in figure
\ref{fig:adaptation}(b) and is referred to as adaptation length.
More specifically, $l_a$ is defined as the maximum boundary-normal
distance from which
\begin{equation}
\mathcal{E}^+(x_2) = |\psi^+(x_2) - \psi^+_{\mathrm{NS}}(x_2)| < \alpha, 
\end{equation}
where the subscript NS denotes variables for no-slip cases, and $\psi$
takes the value of $\langle u_1 \rangle$ or $\langle {u_i'}^2
\rangle^{1/2}$ with $i=1,2,3$, depending on whether the adaptation
length refers to the recovery of the mean velocity profile, or the
streamwise, wall-normal or spanwise r.m.s. velocity fluctuations,
respectively. The thresholding error $\alpha$ is set to be equal to
0.5 plus units for the mean profile and 0.1 for the r.m.s. velocity
fluctuations.  The results show that the flow statistics collapse to
the corresponding no-slip channel above $l_a \sim l$ for the
quantities assessed and various Reynolds numbers. The selected values
of $\alpha$ are admittedly arbitrary and other choices may be
preferred without any major consequences other than a vertical shift
of the adaptation length in figure \ref{fig:adaptation}(b).
%


Then, results from figure \ref{fig:adaptation}(b) can be interpreted
without the need of a virtual wall if the boundary condition is
understood as a flow distortion stirred by the turbulent background
for a depth $l_a^2 \sim \nu_t t$ during a time period of $t \sim
l_a/\langle u_2^2 \rangle^{1/2}$, where $\nu_t$ is the eddy
viscosity. Assuming that $\nu_t \sim \langle u_1 u_2 \rangle / \langle
\partial u_1 /\partial x_2 \rangle$, and that the flow follows
(\ref{eq:slip_bc}) within the adaptation layer defined by $x_2
\lesssim l_a$, then
\begin{equation}
l_a \sim 
\frac{ \langle u_1 u_2 \rangle }{ \langle u_2^2 \rangle^{1/2} 
\left \langle \frac{\partial u_1 }{\partial x_2} \right \rangle } 
\approx 
l \frac{ \left \langle \frac{\partial u_1}{\partial x_2} \frac{\partial u_2 }{\partial x_2} \right \rangle}
{ \left \langle \left( \frac{\partial u_2}{\partial x_2} \right)^2 \right \rangle^{1/2} 
  \left \langle \frac{\partial u_1}{\partial x_2} \right \rangle}, 
\end{equation}
from where it is reasonable to assume that $l_a \sim l$ in first-order
approximation, consistent with the results reported in figure
\ref{fig:adaptation}(b). The existence of this adaptation layer due to
the imposition of an unphysical Robin boundary condition together with
the observations from previous sections suggest that both the
Robin-bounded and wall-bounded channels share identical flow motions
for $x_2>l_a$ once the disturbance by the boundary vanishes.




\subsection{Logarithmic layer without inner-outer scale separation}
\label{subsec:results:log}

\begin{figure}
\begin{center}
 \vspace{0.1cm}
 \psfrag{X}[bc]{$x_2/h$}\psfrag{Y}[cc]{$\eta/h$, $L_\varepsilon/h$}
 \subfloat[]{ \includegraphics[width=0.48\textwidth]{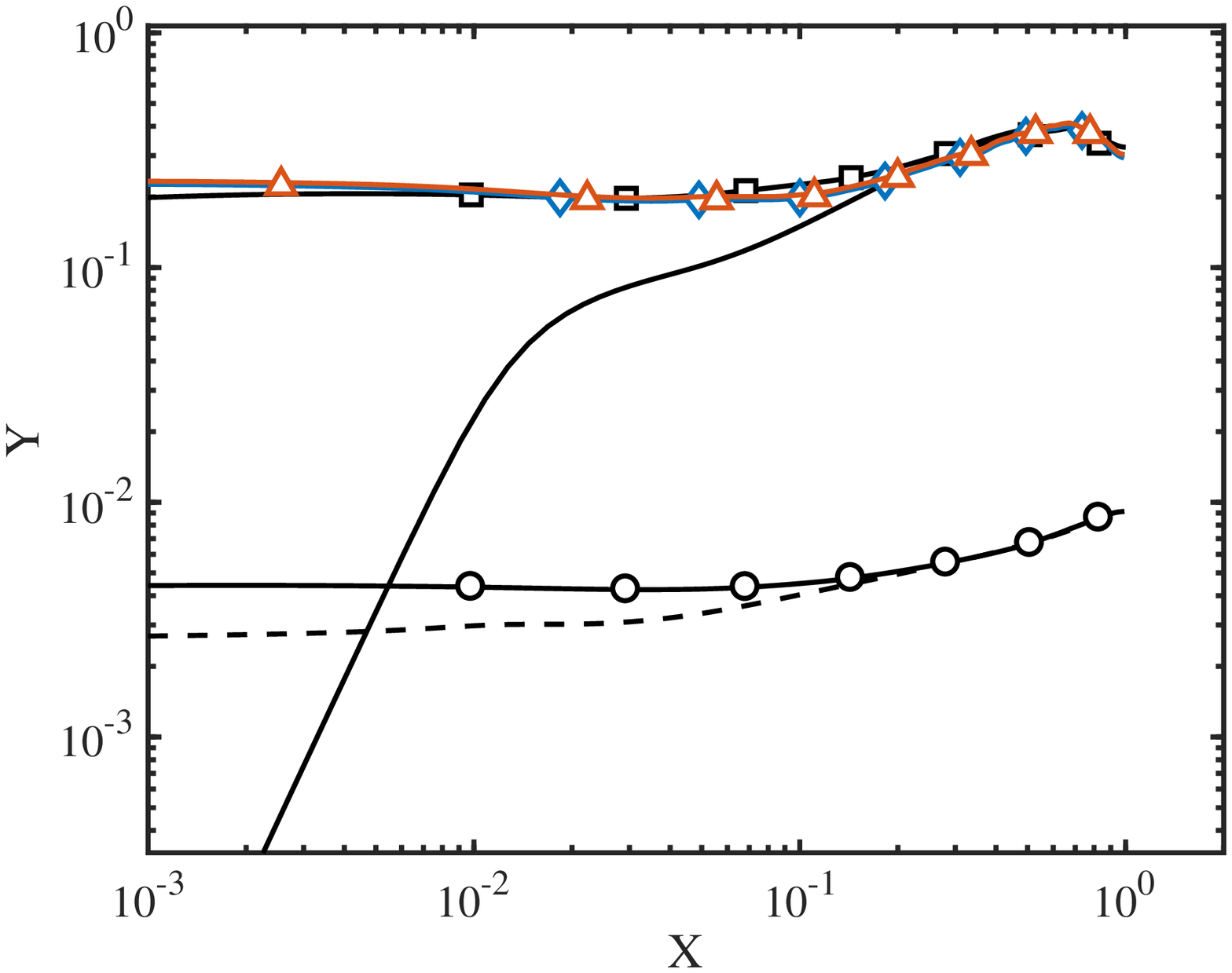} }
 \psfrag{X}[br]{$Re_\tau$}\psfrag{Y}{$\mathcal{E}_l^+$}
 \subfloat[]{ \includegraphics[width=0.475\textwidth]{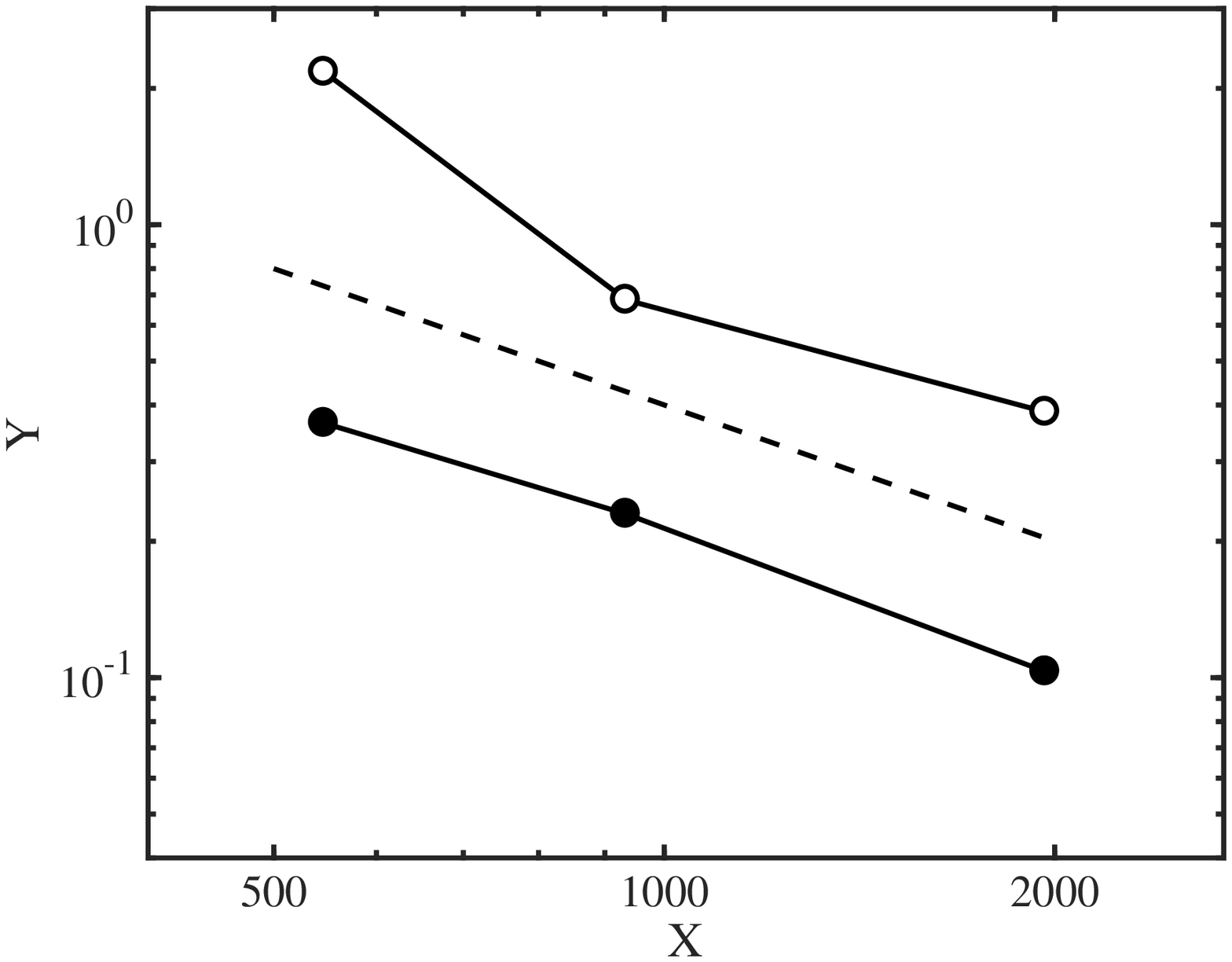} }
 \end{center}
\caption{ (a) Kolmogorov length scale $\eta$ for NS550 (\dashed) and
  R550 ($\circ$), and integral length scale $L_\varepsilon$ for NS550
  (\solid), R550 ($\square$), R950 (\textcolor{blue}{$\Diamond$}), and
  R2000 (\textcolor{red}{$\triangle$}). (b) $\mathcal{E}_l$ as a
  function of $Re_\tau$. Closed and open symbols are for wall-bounded
  and Robin-bounded cases, respectively. The dashed line is
  $\mathcal{E}_l^+ \sim Re_\tau^{-1}$.
\label{fig:log}}
\end{figure}

The Robin boundary condition from (\ref{eq:slip_bc}) imposes a new
length scale to the eddies in the near-wall region. The characteristic
flow length scales of the no-slip and Robin-bounded channels are
plotted in figure \ref{fig:log}(a) as a function of $x_2$. The small
and large scales are represented by the Kolmogorov length scale
$\eta=(\nu^3/\varepsilon)^{1/4}$ and the integral length scale
$L_\varepsilon=(2k/3)^{3/2}/\varepsilon$, respectively, where
$\varepsilon$ is the rate of energy dissipation and $k$ is the
turbulent kinetic energy. Note that $L_\varepsilon$ drops rapidly to
zero as $x_2$ approaches the wall for the no-slip channel, whereas it
remains roughly constant in the Robin-bounded cases. Moreover, the
comparison of $L_\varepsilon$ at three different $Re_\tau$ for the
Robin-bounded channels shows that the integral length scale collapses
in outer units across the entire boundary layer thickness, including
the region close to $x_2/h = 0$. These results can be read as the
disruption of the classic viscous scaling of the active
energy-containing eddies at the wall, i.e., their sizes are a fixed
fraction of $h$ and do not decrease with $Re_\tau$.
%

In spite of the lack of inner-outer layer scale separation with
increasing $Re_\tau$, the mean velocity profile for Robin-bounded
cases (figure \ref{fig:mean}a) tends towards a log-layer as in
wall-bounded channels. This is quantified in figure \ref{fig:log}(b),
which contains the error function
\begin{equation}\label{eq:error_log}
\mathcal{E}_l^+ = \left[ \frac{1}{0.1}
\int_{0.1}^{0.2} \left(
x_2^+ \frac{\partial \langle u_1^+ \rangle} {\partial x_2^+} -
\frac{1}{\kappa} - \frac{x_2}{h} 
\right)^2  \mathrm{d}(x_2/h)
\right]^{1/2},
\end{equation}
with $\kappa = 0.384$ \citep{Lee2015} as a function of
$Re_\tau$. Equation (\ref{eq:error_log}) measures the deviation of
$x_2^+ \partial \langle u_1^+ \rangle / \partial x_2^+$ with respect
to its asymptotic value for a well-developed log-layer, $1/\kappa -
x_2/h$, within the range $x_2/h\in[0.1,0.2]$. The asymptotic value of
$x_2^+ \partial \langle u_1^+ \rangle / \partial x_2^+$ has been
derived by several authors using matched asymptotic expansions
\citep{Mellor1972,Afzal1973,Afzal1976,Phillips1987,Jimenez2007},
\begin{equation}\label{eq:matching}
x_2^+ \frac{\partial \langle u_1^+ \rangle} {\partial x_2^+} \approx
\frac{1}{\kappa} + \frac{x_2}{h} + \frac{\tilde \beta}{Re_\tau},
\end{equation}
with $\tilde \beta$ a Reynolds-number-independent constant.  The
results, plotted in figure \ref{fig:log}(b), show that $\mathcal{E}_l$
is larger for Robin-bounded cases than for wall-bounded cases, but
both set-ups converge to the expected value at a rate close to
$Re_\tau^{-1}$ as predicted by (\ref{eq:matching}). The fact that
Robin-bounded cases approach a logarithmic profile for increasing
$Re_\tau$ without the inner-outer scale separation challenges the
log-layer formulations derived from Millikan's argument
\citep[see][among
  others]{Millikan1938,Wosnik2000,Oberlack2001,Buschmann2003}. Nonetheless,
the Reynolds numbers in the present work are too low to attain a
well-developed log-layer, and therefore, the results are indicative
but not conclusive of the convergence of Robin-bounded cases to an
actual wall-bounded log-layer.
As we are concerned with the outer layer, the analysis above was
performed for a range of wall-normal distances fixed in outer units,
$x_2/h \in[0.1,0.2]$, and the slip length set to a given fraction of
$h$.  The effect of keeping the slip length constant in wall units as
$Re_\tau$ increases is discussed in Appendix \ref{sec:appendixC},
where we show the development of the log-layer in the wall-normal
direction.

\subsection{Wall-attached eddies without walls}\label{subsec:results:attached}

A more detailed analysis of the size of the eddies is provided by the
investigation of three-dimensional regions of the flow where a
quantity of interest is particularly intense. We focus on the regions
of high momentum transfer from \cite{Lozano2012} \citep[see
  also][]{Lozano2014b,Lozano2016}, defined as spatially connected
points in the flow satisfying
\begin{equation}\label{eq:thr:uvs}
-{u_1'} {u_2'} > H \langle {u_1'}^2 \rangle^{1/2} \langle {u_2'}^2 \rangle^{1/2},
\end{equation}
where ${u_1'} {u_2'}$ is the instantaneous point-wise fluctuating
tangential Reynolds stress, and $H$ is a thresholding parameter equal
to $1.75$ obtained from a percolation analysis
\citep{Moisy2004}. Three-dimensional structures are then constructed
by connecting neighbouring grid points fulfilling relation
(\ref{eq:thr:uvs}) and using the 6-connectivity criteria
\citep{Rosenfeld1982}.  The cases under investigation are NS2000 and
R2000, and the total number of structures identified is of the order
of $10^6$. Since we are interested in the outer-layer eddies, the
region $x_2^+ < 100$ was excluded from case NS2000 in order to avoid
spurious contributions from the viscous layer that are not present in
R2000. This is consistent with \cite{Dong2017}, who showed that
objects identified by (\ref{eq:thr:uvs}) are artificially elongated in
the streamwise direction when the region below $x_2^+ = 100$ is
included.

\begin{figure}
\begin{center}
 \vspace{0.1cm}
 \includegraphics[width=0.95\textwidth]{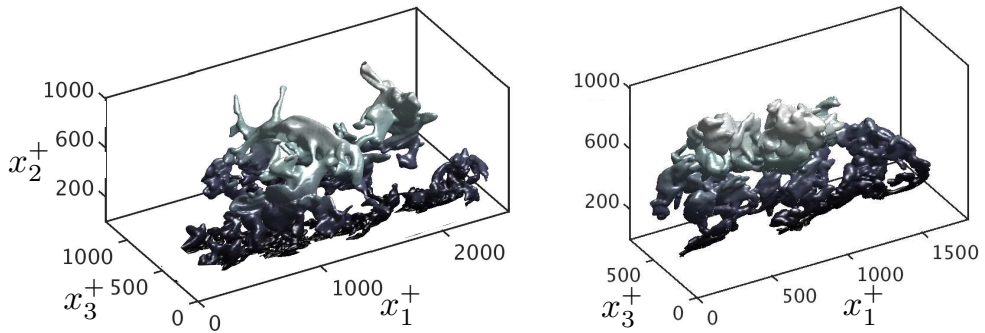}
 \end{center}
\caption{ Example of instantaneous three-dimensional momentum transfer
  structures extracted from R2000 (left) and NS2000 (right). The mean
  flow is from bottom left to top right. Axes are normalised in plus
  units. The colour gradient indicates distance to the wall/boundary.
\label{fig:eddies_examples}}
\end{figure}

Figure \ref{fig:eddies_examples} shows two examples of actual objects
extracted from NS2000 and R2000, and underlines the complex geometries
that may arise. Although visual impressions should not substitute
statistical analysis, the comparison between the two examples
reinforces the idea that the structures in Robin-bounded and wall-bounded
channels are comparatively similar.
%

\begin{figure}
\begin{center}
 \vspace{0.1cm}
 \psfrag{X}{$\Delta_1^+$}\psfrag{Y}{$\Delta_2^+$}
 \subfloat[]{ \includegraphics[width=0.48\textwidth]{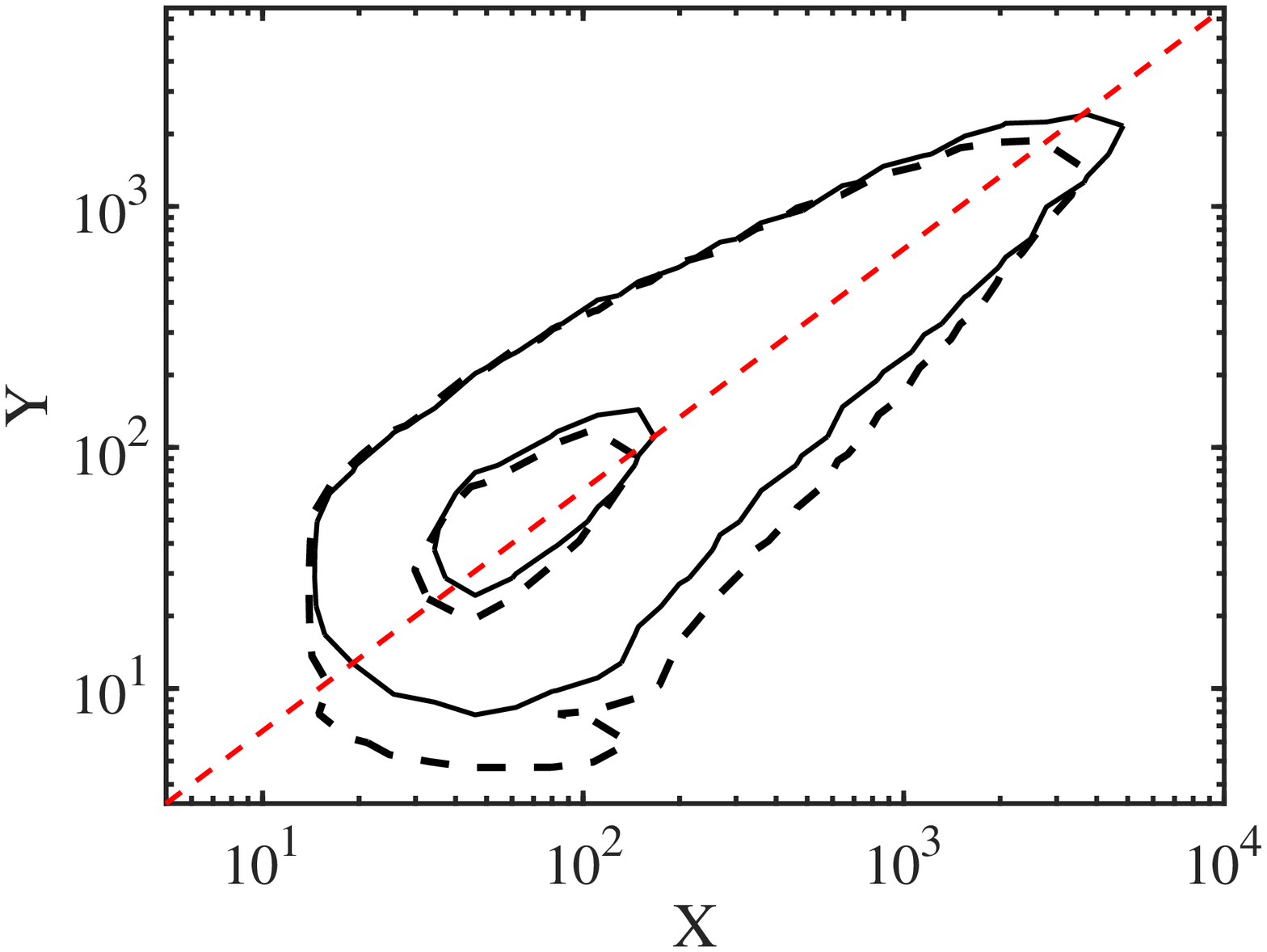} }
 \psfrag{X}{$\Delta_3^+$}\psfrag{Y}{$\Delta_2^+$}
 \subfloat[]{ \includegraphics[width=0.48\textwidth]{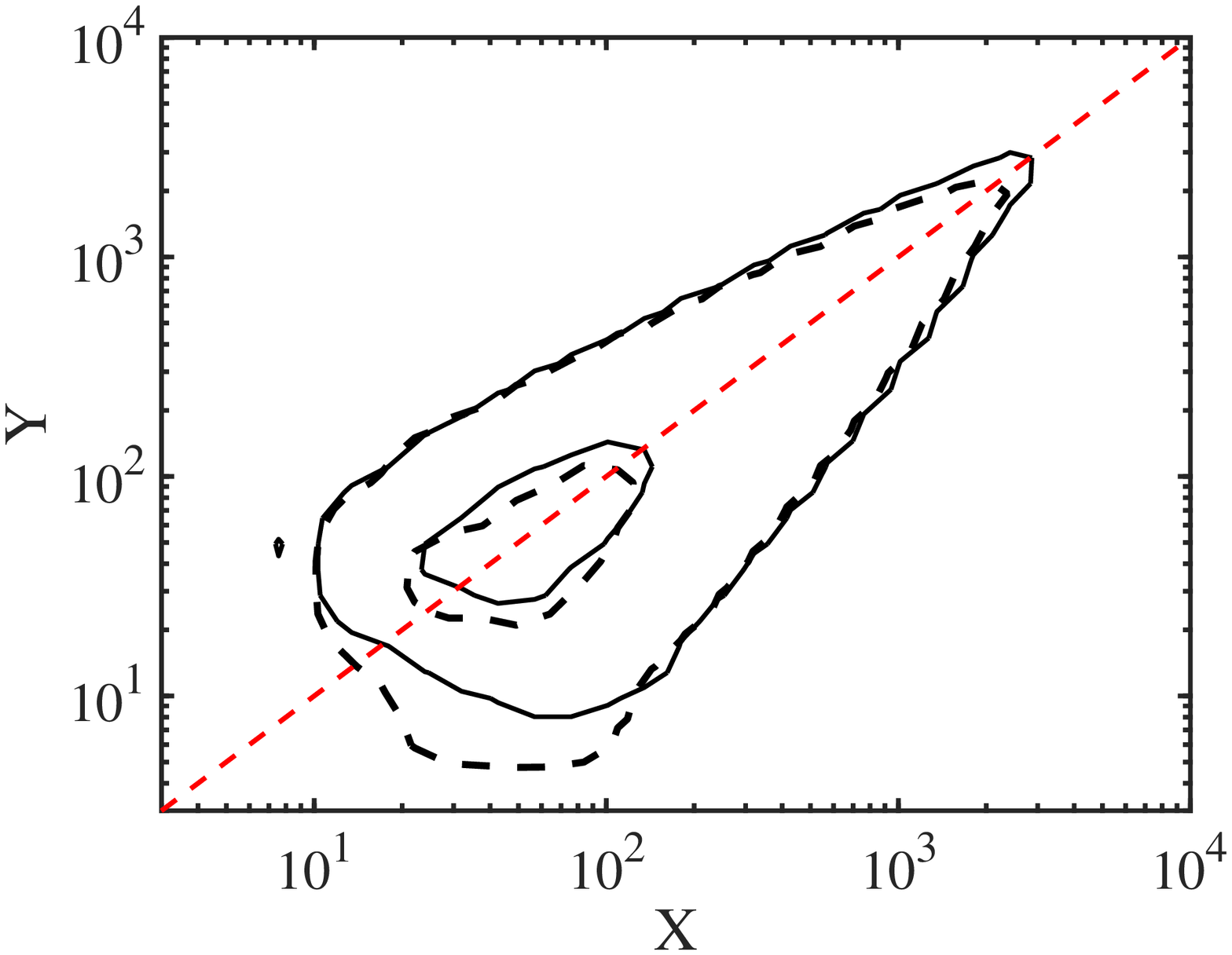} }
 \end{center}
\caption{ Joint probability density functions of the logarithms of the
  boxes sizes of structures, (a) $p(\Delta_1^+,\Delta_2^+)$ and (b)
  $p(\Delta_3^+,\Delta_2^+)$, for NS2000 (\solid) and R2000 (\dashed).
  Contours contain 50\% and 99.8\% of the p.d.f.  The dashed straight
  lines are $\Delta_1 = 1.5\Delta_2$ and $\Delta_1 = \Delta_3$,
  respectively. \label{fig:pdfsizes}}
\end{figure}

\cite{Lozano2012} and \cite{Lozano2014b} showed that objects defined
by (\ref{eq:thr:uvs}) are responsible for most of the momentum
transfer in the log-layer, justifying the choice of intense regions of
$u_1'u_2'$ as representative structures of the flow.  Moreover, these
regions are geometrically and temporally self-similar and can be
considered as sensible contenders for the active wall-attached eddies
envisioned by \citet{Townsend1976}. Therefore, it is interesting to
compare the geometry of these structures in wall-bounded and
Robin-bounded channel flows. The sizes of the objects are measured by
circumscribing each structure within a box aligned to the Cartesian
axes, whose streamwise, wall-normal and spanwise sizes are denoted by
$\Delta_1$, $\Delta_2$ and $\Delta_3$, respectively.  Figure
\ref{fig:pdfsizes} shows the joint probability density functions
(p.d.f.s) of the logarithms of $\Delta_i$, $p(\Delta_1,\Delta_2)$
and $p(\Delta_3,\Delta_2)$, for NS2000 and R2000. Both cases collapse
remarkably well except for small objects close to the boundary, and
follow fairly well-defined linear laws
\begin{equation}
  \Delta_1 \approx 1.5 \Delta_2 \hspace{5mm} \mbox{and} \hspace{5mm}
  \Delta_3 \approx \Delta_2,
  \label{eq:linlaw}
\end{equation}
consistent with the good agreement obtained for the spectra in \S
\ref{subsec:results:stats}. The similarity laws in (\ref{eq:linlaw})
were used in \S \ref{subsec:results:stats} to estimate the sizes of
the eddies at a given $x_2$ location.  The quantitative resemblance of
the momentum eddies in the two flow configurations highlights once
again that the impermeability of the wall (and hence the distance to
the wall) is not a fundamental feature of the stress-carrying motions
in the outer layer of wall-bounded flows.
%

\section{Conclusions}\label{sec:conclusions}


In the present work we have proposed new characteristic velocity,
length, and time scales for the momentum-carrying eddies in the
log-layer of wall-bounded turbulence.  We have hypothesised that the
mean tangential momentum flux and mean shear are the main contributors
to the intensities, lifespan, and sizes of the active
energy-containing motions in the outer region. The proposed
characteristic scales are consistent with the predictions by
Townsend's attached eddy model and extend its applicability to flows
with different mean momentum flux. 

The mechanism proposed is as follows. The mean tangential momentum
transfer defines a characteristic velocity scale $u^*$ at each
wall-normal distance. The role of $u^*$ is twofold: it controls the
intensities of the active eddies and the mean shear. The size of the
eddies is governed by the length scale $l^*$ defined in terms of $u^*$
and the characteristic time scaled $t^*$ imposed by the mean shear. In
this framework, the no-slip and impermeability constraints of the wall
are not directly involved in the organisation of the outer flow, and
the role of the wall is relegated to serve as a proxy to sustain the
mean momentum flux. The scaling proposed has been successfully
assessed through a set of idealised numerical studies in channel flows
with $x_2$-dependent body forces and modified streamwise velocity
profiles.

We have further addressed the question of whether the impermeability
of the wall is a foundational component of the outer-layer of wall
turbulence by designing a new numerical experiment where the walls of
the channel are replaced by a Robin boundary condition. In the
resulting flow, instantaneous wall transpiration is allowed for scales
comparable to the size of the log-layer motions to the extent that the
wall-normal distance can no longer be a relevant length scale. We have
referred to this configuration as Robin-bounded channel flow as
opposed to the traditional wall-bounded channel.

A detailed inspection of the one-point statistics, spectra, and
three-dimensional structures responsible for the momentum transfer has
shown that both wall-bounded and Robin-bounded channel flows share
identical outer-layer motions, and we have interpreted this evidence
as an indication that the same physical processes occur in both flow
configurations. The results are consistent with previous studies on
rough walls and idealised numerical experiments with modified walls,
although it is important to stress that in the present set-up wall
transpiration is allowed for length scales of the order of the
log-layer eddies. In that sense, our findings generalise the
Townsend's similarity hypothesis for permeable boundaries, and
reinforce the conclusion from previous studies which highlighted the
secondary role played by the wall.

\section*{Acknowledgements}

This work was supported by NASA under Grant \#NNX15AU93A and by ONR
under Grant \#N00014-16-S-BA10.  The authors thank Prof. Parviz Moin,
Prof. Javier Jim\'enez, Dr. Perry Johnson, and Dr. Minjeong Cho for
their insightful comments on previous versions of this manuscript.

\appendix

\section{Characteristic velocity without wall-normal coordinate}\label{sec:appendixB}

\begin{figure}
 \begin{center}  
   \vspace{0.1cm}
   \psfrag{X}{$x_2/h$}\psfrag{Y}{$\langle {u_i'}^{2*} \rangle^{1/2}$}
   \includegraphics[width=0.48\textwidth]{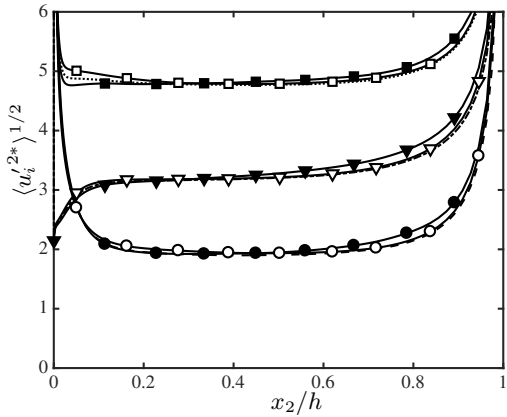}
 \end{center}
\caption{ Streamwise (circles and \dashed), wall-normal (triangles and
  \dchndot), and spanwise (squares and \dotted) root-mean-squared
  velocity fluctuations scaled with $u^*$. Symbols are for NS550-p
  (open) and NS550-n (closed). Lines without symbols are for
  NS550. For clarity, the profiles for the wall-normal and spanwise
  root-mean-squared velocity fluctuations are shifted vertically by
  2.0 and 3.4 wall units. \label{fig:rms_eps_ver2}}
\end{figure}

The factor $\sqrt{1-x_2/h}$ in the definition of $u^\star$ serves only
a practical purpose, i.e., $u^\star=u_\tau$ for traditional turbulent
channel flows. Using $u^*$ instead of $u^\star$ does not degrade the
quality of the scaling reported in \S \ref{subsec:results:flux} except
for a region very close or very far from the wall where the scaling is
no longer applicable owing to the presence of viscous effects or the
lack of mean shear. Figure \ref{fig:rms_eps_ver2} (analogous to figure
\ref{fig:rms_eps}d) shows the scaling of the r.m.s. fluctuating
velocities using as characteristic velocity scale $u^*$. The results
show that the r.m.s. velocity fluctuations collapse and remain roughly
constant across the layer defined by $0.1h<x_2<0.8h$ for the cases
investigated.
%

\section{Flow structure at the boundary for channels with Robin boundary condition}\label{sec:appendixA}

\begin{figure}
\begin{center}
\vspace{0.1cm}
\includegraphics[width=1.05\textwidth]{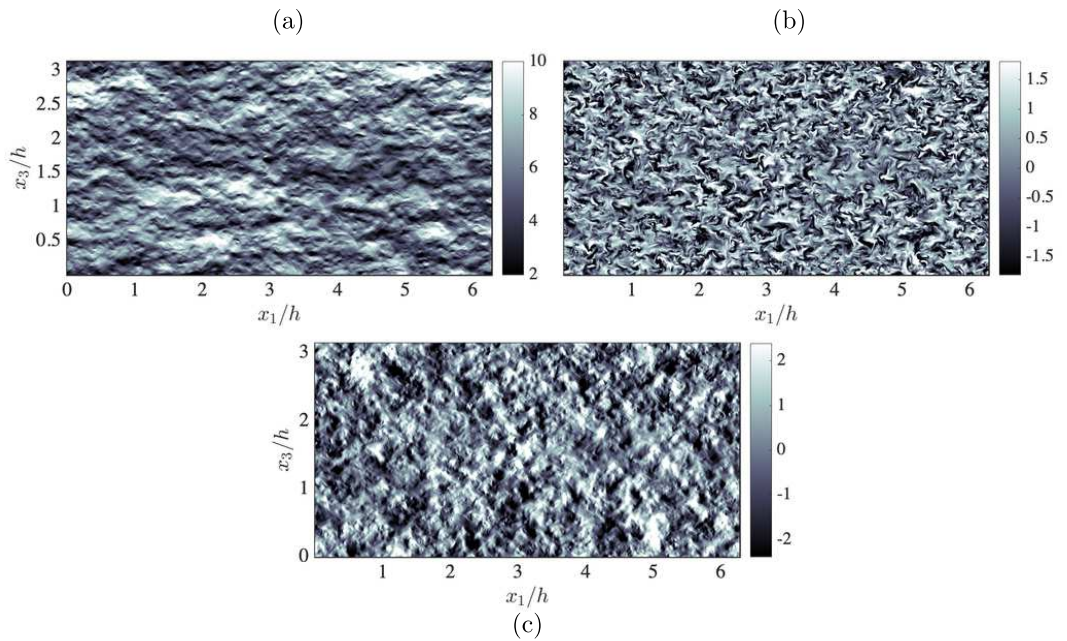}
\end{center}
\caption{Instantaneous $x_1$--$x_3$ planes of the (a) streamwise, (b)
  wall-normal, and (c) spanwise velocity for R2000 at $x_2/h=0$. The
  mean flow is from left to right. The colour bars show velocity
  normalised in wall units.\label{fig:snapshots_slip_wall}}
\end{figure}

\begin{figure}
\begin{center}
 \vspace{0.1cm}
 \psfrag{X}{$l/h$}\psfrag{Y}{$\left. \langle {u_i'}^{2+} \rangle^{1/2} \right|_w$}
 \subfloat[]{\includegraphics[width=0.45\textwidth]{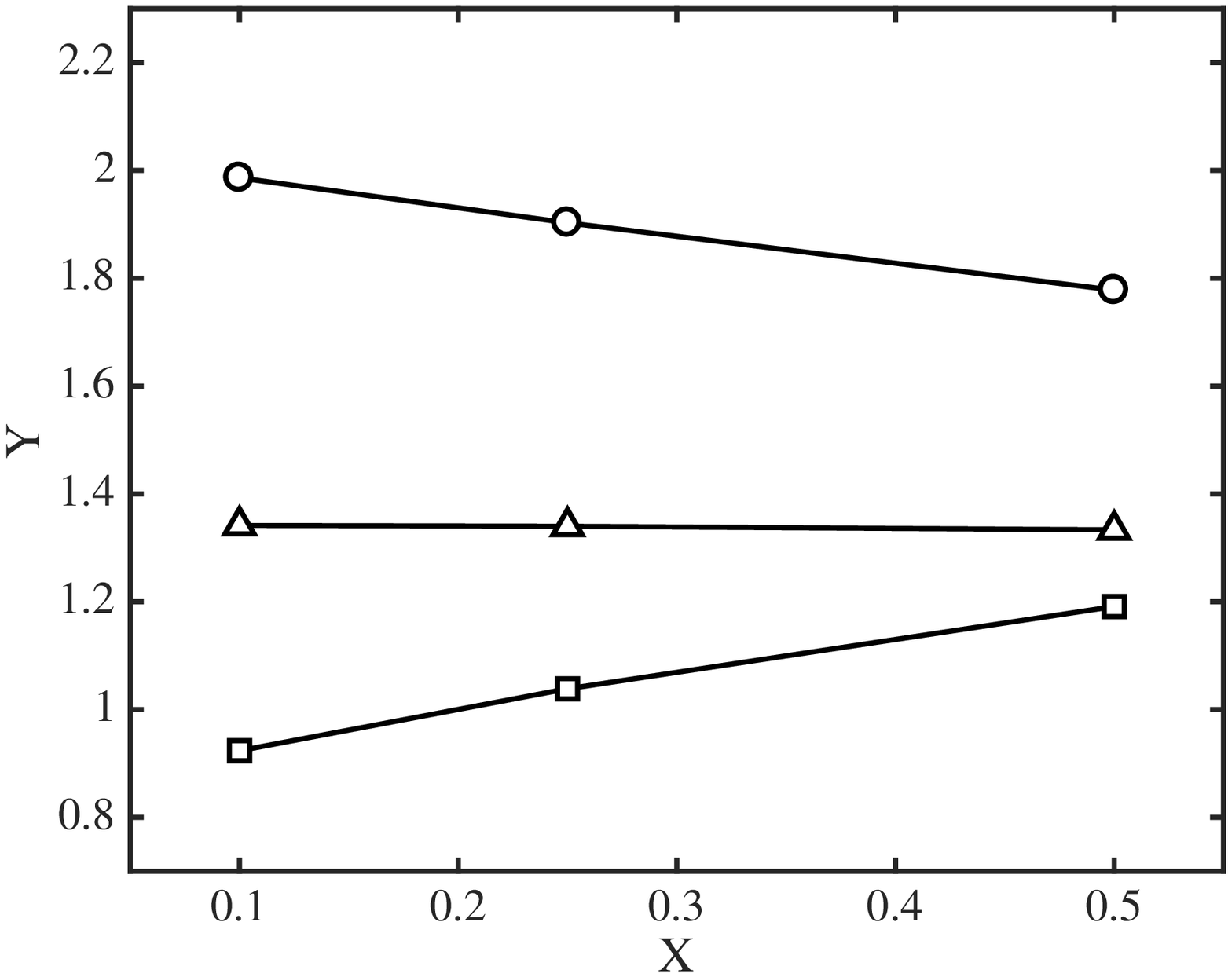} }
 \hspace{0.1cm} 
 \psfrag{X}{$Re_\tau$}\psfrag{Y}{$\left. \langle {u_i'}^{2+} \rangle^{1/2} \right|_w$}
 \subfloat[]{\includegraphics[width=0.47\textwidth]{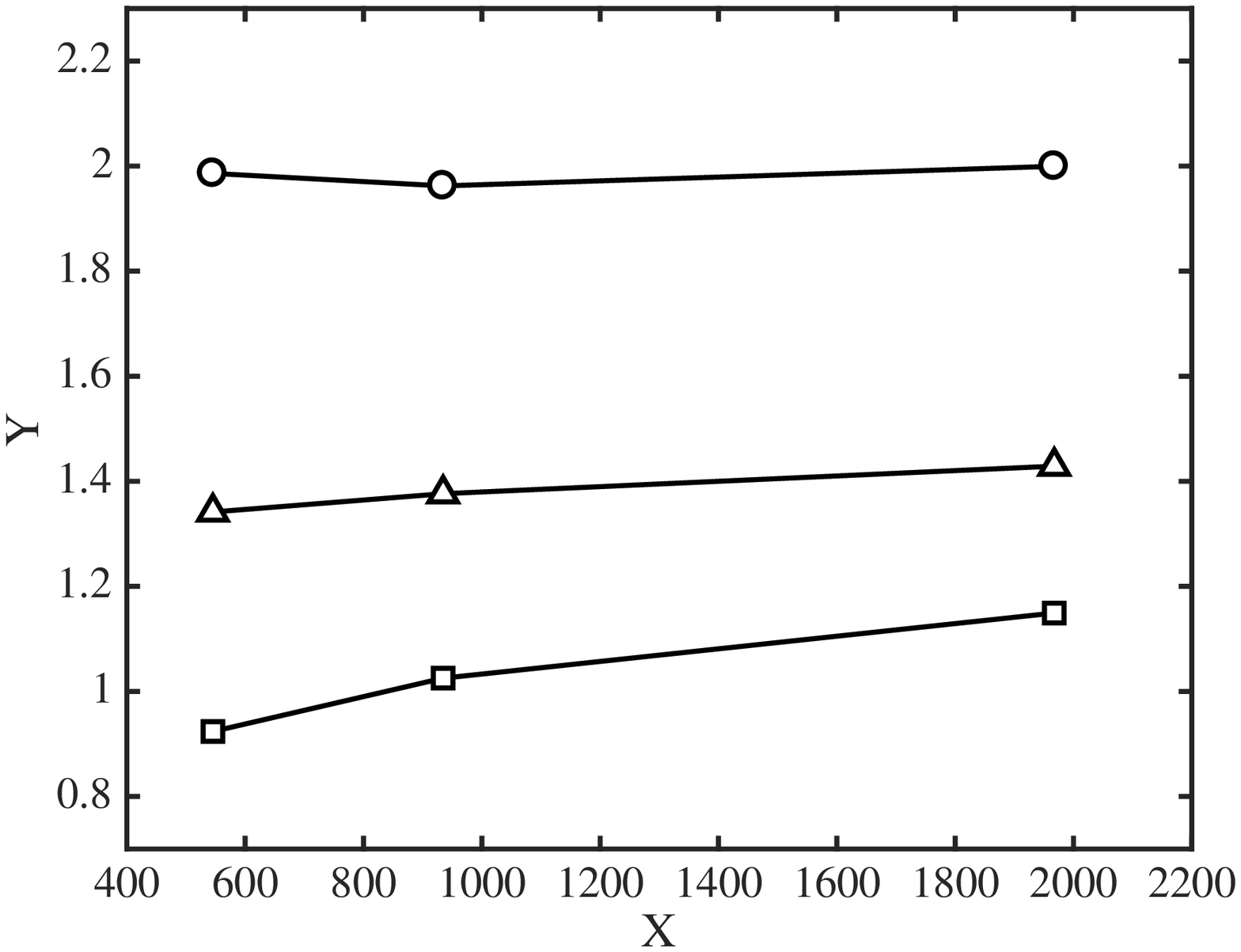} }
\end{center}
\caption{ Streamwise ($\circ$), wall-normal ($\square$), and spanwise
  ($\triangle$) root-mean-squared fluctuating velocities at
  $x_2/h=0$ for Robin-bounded cases as a function of (a) slip length
  and (b) friction Reynolds number. \label{fig:rms_wall}}
\end{figure}

In the present appendix, we document the intensities and sizes of the
velocity structures at $x_2/h=0$ for turbulent channels with Robin
boundary conditions. Snapshots of the three instantaneous velocities
are shown in figure \ref{fig:snapshots_slip_wall} to provide a
qualitative assessment of the characteristic structure of the flow at
the boundary. The flow exhibits an elongated structure for the
streamwise velocity reminiscent of the traditional near-wall streaks,
but with streamwise sizes much shorter than those reported for no-slip
channels in the region $x_2^+<10$. The wall-normal and spanwise
velocities show a quasi-isotropic organisation in the wall-parallel
plane.
%
%
The intensities of the three fluctuating velocities at $x_2/h=0$ are
quantified in figure \ref{fig:rms_wall} as a function of the slip
length and Reynolds number. Intensities remain fairly constant with
$Re_\tau$ and change slowly with the slip length. For increasing slip
length, the streamwise r.m.s. velocity decreases, whereas the
wall-normal r.m.s. velocity increases.  The spanwise r.m.s. velocity
is roughly constant for the range of slip lengths studied. The trend
suggests that the three velocities converge to an isotropic state for
increasing slip length, which may be caused by the fact that the Robin
boundary condition resembles free slip $\partial u_i/\partial x_2
\approx 0$ in such a limit. Note that the free-slip limit is not of
interest in the present work as the flow is unable to support a mean
shear.
%

\begin{figure}
\begin{center}
 \vspace{0.5cm}
 \includegraphics[width=0.8\textwidth]{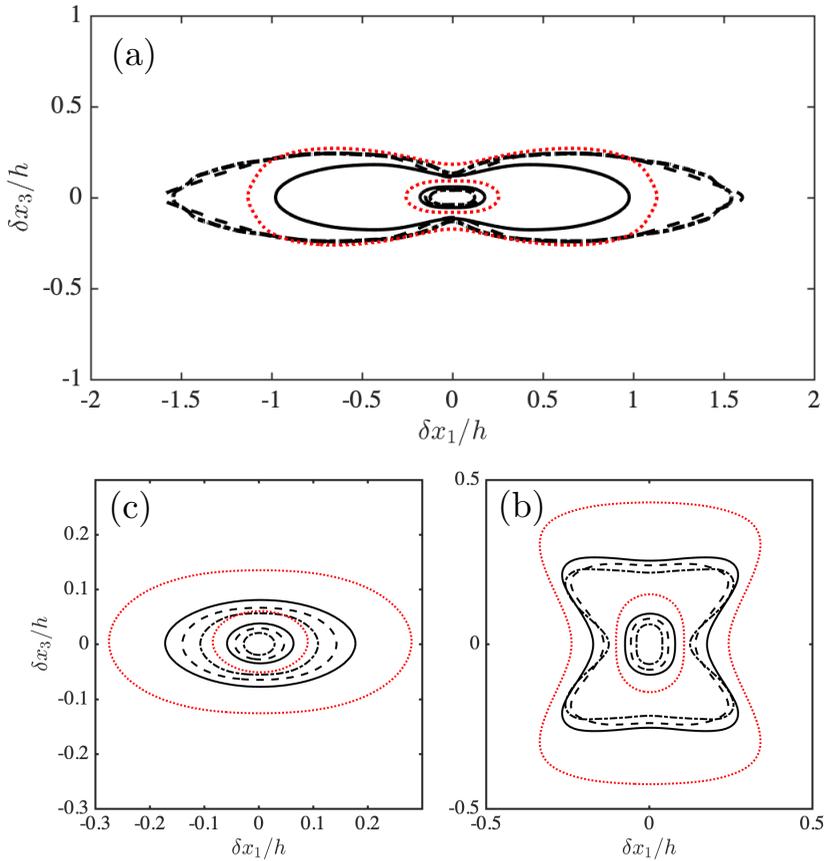}
\end{center}
\caption{ Wall-parallel ($x_1-x_3$) correlations of the (a)
  streamwise, (b) wall-normal, and (c) spanwise velocity for R550
  (\solid), R950 (\dashed), R2000 (\dchndot), and R550-l1
  (\textcolor{red}{\dotted}). Contours are 0.05 and 0.5 of the maximum
  value.  The mean flow is from left to right.
\label{correlation:rms_wall}}
\end{figure}

The sizes of the velocity structures are quantified in figure
\ref{correlation:rms_wall} through the two-point correlation of the
velocities as a function of the streamwise and spanwise increments
$\delta x_1$ and $\delta x_3$, respectively. For a constant slip
length, the iso-contours of the correlation scale reasonably well
using $h$ for $u_1$ and $u_3$ and $Re_\tau>550$. Some dependence on
the Reynolds number can be observed for the correlation of the
wall-normal velocity, although the scaling with $h$ is still superior
than the scaling using wall units (not shown).  Increasing the slip
length while maintaining $Re_\tau$ (dotted-red line in figure
\ref{correlation:rms_wall}) shortens the streamwise velocity
correlation and enlarges the size of the $u_2$ and $u_3$ structures.
%

\section{Logarithmic layer in Robin-bounded cases with $l$ fixed in wall units}\label{sec:appendixC}
\begin{figure}
\begin{center}
 \vspace{0.1cm}
 \psfrag{X}{$x_2^+$}\psfrag{Y}{$\Xi^+$}
 \subfloat[]{\includegraphics[width=0.45\textwidth]{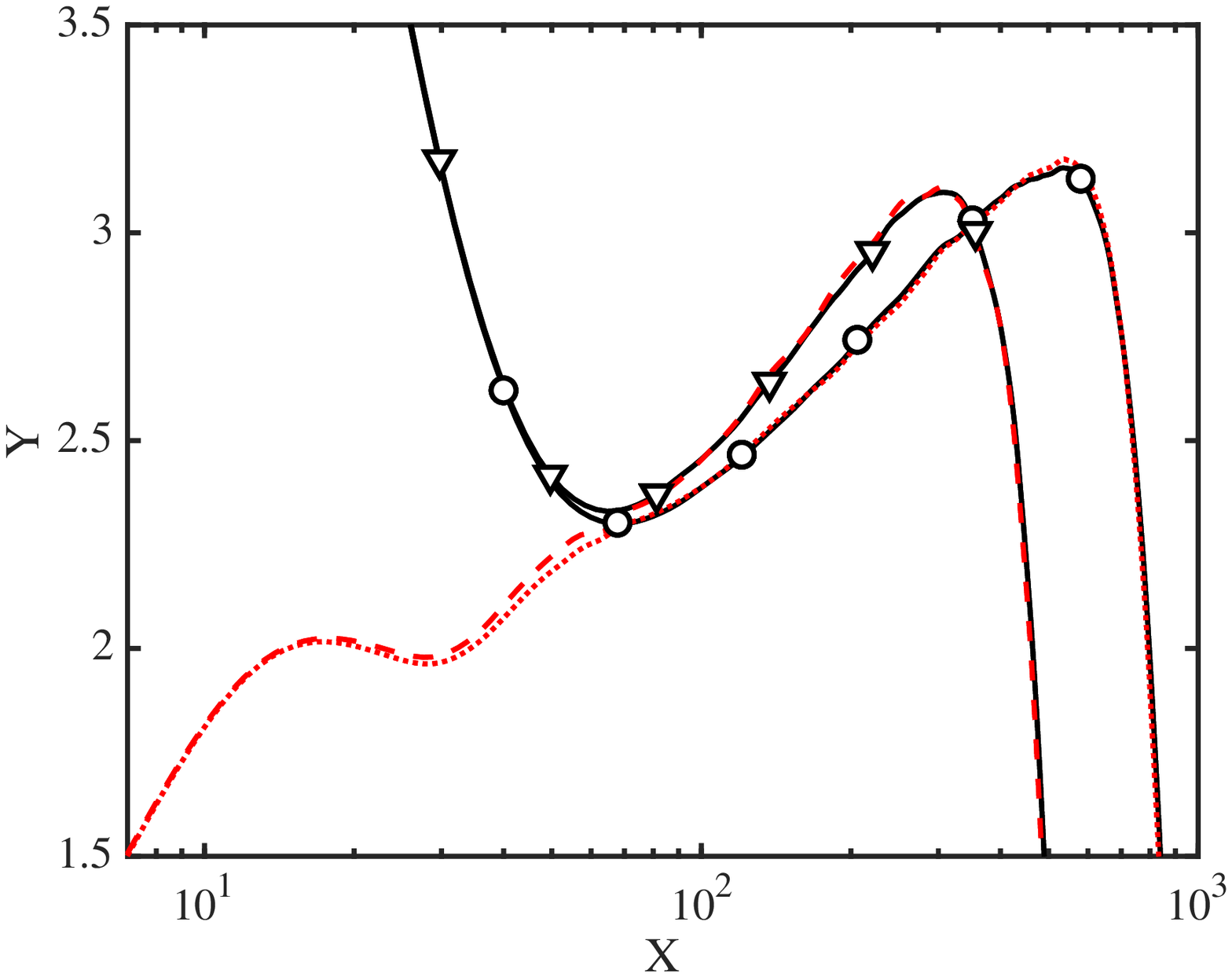} }
 \hspace{0.1cm} 
 \psfrag{X}{$Re_\tau$}\psfrag{Y}{$\mathcal{E}_{l2}^+$}
 \subfloat[]{\includegraphics[width=0.45\textwidth]{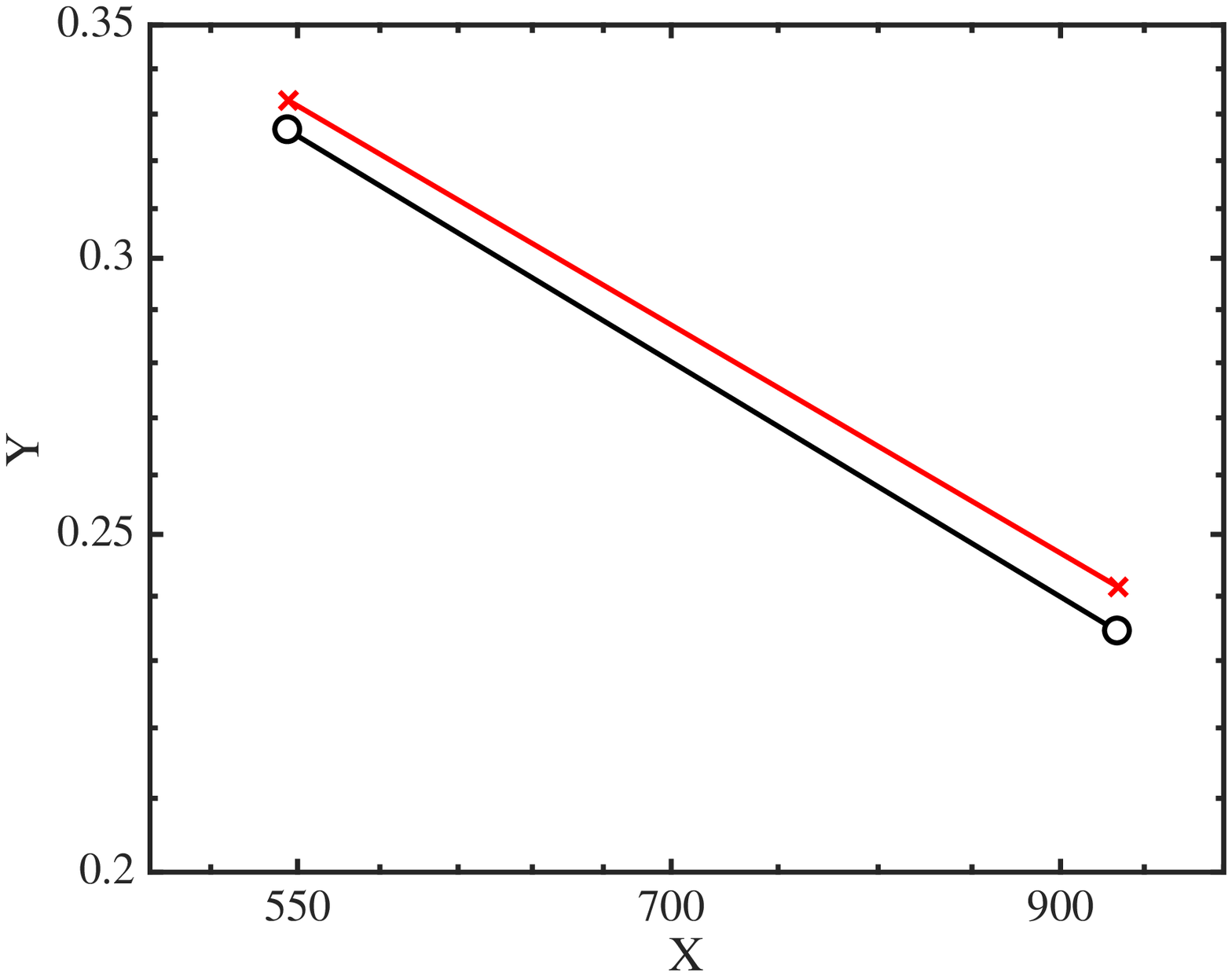} }
\end{center}
\caption{ (a) Diagnostic function $\Xi = x_2 \partial \langle u_1
  \rangle /\partial x_2$ for NS550 ($\triangledown$), NS950 ($\circ$),
  R550-lplus (\textcolor{red}{\dashed}), and R950-lplus
  (\textcolor{red}{\dotted}). (b) $\mathcal{E}_{l2}$ as a function of
  $Re_\tau$ for wall-bounded cases NS550 and NS950 ($\circ$), and
  Robin-bounded cases R550-lplus and R950-lplus
  (\textcolor{red}{$\times$}).
\label{fig:log_l_20plus}}
\end{figure}

%
We perform two additional DNSs of Robin-bounded channels with slip
length $l^+=20$ at $Re_\tau\approx 550$ (labelled as R550-lplus) and
$Re_\tau\approx 950$ (labelled as R950-lplus). The numerical set-up is
identical to that discussed in \S \ref{subsec:numerical:slip} for
cases R550 and R950. Figure \ref{fig:log_l_20plus}(a) shows the
diagnostic function $\Xi = x_2 \partial \langle u_1 \rangle /\partial
x_2$ for Robin-bounded cases compared to their no-slip equivalents
NS550 and NS950.  The agreement between the Robin-bounded and
wall-bounded cases is excellent for $x_2^+ \gtrsim 70$. The results
are consistent with the adaptation-length argument presented in \S
\ref{subsec:results:adaptation}: fixing $l$ in wall units introduces a
perturbation which propagates in the wall-normal direction for a
distance $\mathcal{O}(\nu/u_\tau)$.  While the Reynolds numbers above
are too low to develop a well-defined log-layer, the collapse of the
diagnostic functions reported in figure \ref{fig:log_l_20plus}(a)
suggests the emerging of a log-layer for both wall-bounded and
Robin-bounded cases.  The deviation from the log-layer is further
quantified by the error function (see \S \ref{subsec:results:log})
\begin{equation}
\mathcal{E}_{l2}^+ = \left[ \frac{1}{0.2 - 100/Re_\tau}
\int_{100/Re_\tau }^{0.2}
\left(
x_2^+ \frac{\partial \langle u_1^+ \rangle} {\partial x_2^+} -
\frac{1}{\kappa} - \frac{x_2}{h} 
\right)^2  
\mathrm{d}(x_2/h)
\right]^{1/2},
\end{equation}
where the lower integration limit is chosen in wall units as expected
for the near-wall edge of the log-layer.  The results, included in
figure \ref{fig:log_l_20plus}(b), show that Robin-bounded cases with
$l^+=20$ tend to a log-layer for $x_2/h\in[100/Re_\tau ,0.2]$ with a
similar error and at a similar rate as wall-bounded cases. In the
cases presented here, the inner-outer scale separation, defined by
$h/l$, increases with $Re_\tau$. Thus the development of a log-layer
may be anticipated by invoking Millikan's argument.
%


\bibliographystyle{jfm}
\bibliography{wall_less}

\end{document}